\documentclass[12pt]{article}
\usepackage{authblk}
\usepackage{amsmath,amssymb}
\usepackage{setspace,subfig}
\usepackage{natbib}
\usepackage{longtable}
\usepackage{relsize}
\renewcommand{\harvardurl}[1]{\textbf{URL:} \url{#1}}
\usepackage{booktabs}
\usepackage{pgf}
\usepackage{multirow}
\usepackage{setspace}
\usepackage{times}
\usepackage{mathtools}
\usepackage{tikz}
\usetikzlibrary{arrows,shapes.arrows,shapes.geometric,
	backgrounds,decorations.pathmorphing,positioning,fit,automata,
	shapes.swigs} %% This requires that pgflibraryshapes.swigs.code.tex
\tikzset{
	>=stealth',
	true/.style={
		rectangle,
		draw=black, very thick,
		text width=6.5em,
		minimum height=2em,
		text centered,
		fill=gray, opacity = 0.5},
	punkt/.style={
		rectangle,
		rounded corners,
		draw=black, very thick,
		text width=6.5em,
		minimum height=2em,
		text centered},
	est/.style={
		circle,
		draw=black, very thick,
		text centered},
	shade/.style={
		circle,
		draw=black, very thick, fill=gray!50,
		text centered},
	weight/.style={
		circle,
		draw=black, very thick,
		text width=6.5em,
		minimum height=2em,
		text centered},
	pil/.style={
		->,
		thick,
		shorten <=2pt,
		shorten >=2pt,},
	double/.style={
		<->,
		thick,
		shorten <=2pt,
		shorten >=2pt,},
	dash/.style={
		dashed,
		thick,
		shorten <=2pt,
		shorten >=2pt,},
	dashdouble/.style={
		<->,
		dashed,
		thick,
		shorten <=2pt,
		shorten >=2pt,}
}
\usepackage{enumerate}

\usepackage{booktabs}
\usepackage{amsfonts}
\usepackage[latin1]{inputenc}
\usepackage{soul,color}
\usepackage{chngcntr}
\usepackage{verbatim}
\usepackage{changepage}
\usepackage{amssymb}
\usepackage{pgf}
\newcommand{\bm}[1]{\mbox{\boldmath{$#1$}}}

\usepackage{caption} 
\usepackage[left=1in,right=1in,top=1in,bottom=1in]{geometry}

\usepackage{xfrac}
\usepackage{graphicx}
\newcommand{\ind}{\rotatebox[origin=c]{90}{$\models$}}

\newtheorem{theorem}{Theorem}

\newtheorem{remark}{Remark}
\usepackage[T1]{fontenc}

\newtheorem{proof}{Proof}
\allowdisplaybreaks
\newcommand\numberthis{\addtocounter{equation}{1}\tag{\theequation}}

\usepackage{algorithm}
\date{}
\usepackage{tikz}

\usepackage{soulpos}
\soulregister\cite7
\soulregister\citep7
\soulregister\ref7
\soulregister\emph7
\soulregister\eqref7
\soulregister\pageref7
\definecolor{mygreen}{RGB}{144,241,47}
\newcommand{\hlc}{}

\newcommand{\LATE}{\delta^{L}}
\newcommand{\MLATE}{\delta^{M}}

\newcommand{\m}{\mathcal{M}}

\def\bSig\mathbf{\Sigma}

\usepackage[figuresright]{rotating}

	{\raise0.5ex\hbox{$#1$}\! \left/ \! \lower0.5ex\hbox{$#2$}\right.}
	
\begin{document}

		\markboth{L. Wang, Y. Zhang, T.S. Richardson and J.M. Robins}{Estimation of local treatment effects}
	
	%% Here are the title, author names and addresses

 \centerline{\large\bf Estimation of local treatment effects under the binary instrumental variable model}
% \vspace{2pt}
%  \centerline{\large\bf under the binary instrumental variable model}
\vspace{.25cm}
\vspace{.4cm} \centerline{$^{1}$Linbo Wang, $^{2}$Yuexia Zhang, $^{3}$Thomas S. Richardson, $^{4}$James M. Robins} \vspace{.4cm} \centerline{\it $^{1,2}$University of Toronto, $^{3}$University of Washington, $^{4}$Harvard T.H. Chan School of Public Health
} \vspace{.55cm} 
\par

	\begin{abstract}
	Instrumental variables are widely used to deal with unmeasured confounding in observational studies and imperfect randomized controlled trials. In these studies, researchers often target the so-called local average treatment effect as it is identifiable under mild conditions. In this paper, we consider estimation of the local average treatment effect under the binary instrumental variable model. We discuss the challenges for causal estimation with a binary outcome, and show that surprisingly, it can be more difficult than the case with a continuous outcome. We propose novel modeling and estimating procedures that improve upon existing proposals in terms of model congeniality, interpretability, robustness or efficiency. Our approach is illustrated via simulation studies and a real data analysis.
	\end{abstract}
	

\section{Introduction}

Unmeasured confounding is a common threat to draw valid causal inference in practice.  It can occur  in observational studies as well as imperfect randomized controlled studies where participants may not comply with the assigned treatment. Instrumental variable methods which seek to address this, are widely used in economics, biostatistics and epidemiology to estimate causal effects when unmeasured confounders may be present.  Intuitively, an instrumental variable is a pre-treatment covariate that is associated with the outcome only through its effect on the treatment. In practice, the condition above is often reasonable only after controlling for a set of baseline covariates. 

Traditionally, instrumental variable methods have aimed to estimate average treatment effects \citep{wright1928tariff,goldberger1972structural}. Identification of the average treatment effects, however, relies on untestable homogeneity assumptions involving unmeasured confounders  \citep[e.g.][]{hernan2006instruments,wang2018bounded}. An alternative proposed by \cite{imbens1994identification} and \cite{angrist1996identification} is to estimate  the so-called local average treatment effects   as they can be non-parametrically identified under a certain monotonicity assumption.  
% In the non-compliance setting, the local average treatment effect is also known as the complier average causal effect, since it concerns the treatment effect in a so-called compliers subgroup that would comply with the treatment assignment irrespective of what that assignment is.
%\change{(Why the LATE might be of interest in practice?)}
In the non-compliance setting, the local average treatment effects may be of interest in practice 
since if the local effect indicates that
treatment is advantageous then this can be used as an argument for increasing the incentives for taking the treatment.

The estimation problem of local average treatment effects has been studied extensively for continuous outcomes  \citep[e.g.][]{abadie2002instrumental,abadie2003semiparametric,tan2006regression,okui2012doubly,ogburn2015doubly}.  However, as explained in detail in Section \ref{sec:framework}, 
direct application of these methods to binary outcomes is often inappropriate. 
%For example, with a binary outcome, the posited models in \cite{okui2012doubly} and \cite{ogburn2015doubly}'s doubly robust estimating approaches  are variation dependent of each other and hence can be uncongenial \citep{meng1994multiple}, while  the  semiparametric regression estimator of \cite{tan2006regression} may fall outside of  the parameter space.  Although the weighted estimating equation approach of \cite{abadie2003semiparametric} can be adapted to estimate the local average treatment effect with a binary outcome by specifying a separate logistic  model for the so-called local average response function within each treatment arm,    the logistic models do not offer direct insight into which baseline variables are important for the causal effect of interest,  nor do they allow for doubly robust estimation of the causal parameters \citep{richardson2017modeling}. 
 Furthermore, with the exception of \cite{ogburn2015doubly} and \cite{abadie2003semiparametric}, most existing methods only focus on the additive local average treatment effect but not the multiplicative local average treatment effect; the latter is of common interest with binary outcomes as it measures the  causal effect on the relative risk scale.   In related work, the  Wald-type estimator of \cite{didelez2010assumptions}  can be shown to be approximately equal to the multiplicative local average treatment effect under a monotonicity assumption \citep{clarke2012instrumental}.
%  \cite{wang2018bounded} studied a different problem of estimating the average treatment effect, but their regression-based estimator can be shown to be consistent for the local average treatment effect under a monotonicity assumption, without suffering  from the aforementioned problems. However,  their approach cannot be adapted to estimate the multiplicative local average treatment effect. Furthermore, as we explain later in Remark \ref{remark:optimal}, even for the local average treatment effect, in general their estimator fails to achieve the semiparametric efficiency bound.
%  their approach does not give rise to the efficient doubly robust g-estimator. 

  In this paper, we propose novel estimating procedures for both the additive and multiplicative local average treatment effects with a binary outcome that (a) ensure the posited models are variation independent and hence congenial  with each other \citep{meng1994multiple}; (b) ensure the resulting estimates lie in the natural non-trivial parameter space;  (c) directly parameterize  the local average treatment effect curves to improve interpretability and reduce the risk of model mis-specification \citep{ogburn2015doubly}; (d) allow for efficient and truly doubly robust estimation of the causal parameter of interest.  \hlc{To the best of 
 our knowledge, for the additive local average treatment effect, our procedure is the first one that achieves (d); for the multiplicative local average treatment effect, our procedure is the first one that achieves (a),(c) or (d); see also Remark \ref{remark:simple}. }

\section{Framework, notation and existing estimators}
\label{sec:framework}

Consider the problem of causal effect estimation with  a binary exposure indicator $D$ and a binary outcome  $Y$.  Suppose  the effect of $D$ on $Y$ is subject to confounding by observed variables $X$ as well as  unobserved variables $U$. Following the potential outcome framework, we assume $D(z)$, the potential exposure if the instrumental variable would take value $z$, to be
well-defined. Similarly, we assume $Y(z,d)$, the outcome that would have been observed if a unit were exposed to $d$ and the instrument had taken value $z$,  to be well-defined.
We assume we also observe a binary instrumental variable $Z$ that satisfies the following assumptions \citep{angrist1996identification}:
\begin{itemize}
	\item[A1] Exclusion restriction: for all $z$ and $z^\prime$, $Y(z,d) = Y(z^\prime,d)\equiv Y(d)$ almost surely;
	\item[A2] Independence:  $Z \ind (Y(d),D(z)) \mid X, d=0,1, z=0,1$;
	\item[A3] Instrumental variable relevance: $P(D(1) = 1\mid X) \neq P(D(0)=1\mid X)$ almost surely;
	\item [A4] Positivity: there exists $\sigma>0$ such that $\sigma < P(Z=1\mid X) < 1-\sigma$ almost surely;
	\item[A5] Monotonicity: $D(1) \geq D(0)$ almost surely. 
\end{itemize}

Notably implicit in the notation $D(z)$ is that the instrument $Z$ is causal so that A3 implies that $Z$ has a non-zero causal effect on $D$. 
  Figure \ref{DAG:iv_model} gives a simple illustration of the
conditional instrumental variable model; see Figure S2 in the Supplementary Material for another example.

\begin{figure}[!htbp]
		\begin{tikzpicture}[->,>=stealth',node distance=1cm,pre/.style={->,>=stealth,ultra thick,black,line width = 1.5pt}, 
		swig vsplit={gap=5pt, line color right=red}]] %% Set style for split nodes
		%nodes
		\begin{scope}
		\node[est] (Z) {$Z$};
		\node[est, right = of Z] (D) {$D$};
		\node[est, right = of D] (Y) {$Y$};
		\node[shade, below = of D] (U) {$U$};
		\node[est, above = of D] (X) {$X$};
		\path[pil] (Z) edgenode {} (D);
		\path[pil] (D) edgenode {} (Y);
		\path[pil] (U) edgenode {} (D);
		\path[pil] (U) edgenode {} (Y);
		\path[pil] (X) edgenode {} (Z);
		\path[pil] (X) edgenode {} (D);
		\path[pil] (X) edgenode {} (Y);
		\node[name=la, below=2cm of D]{(a). A Directed Acyclic Graph.};
		\end{scope}
%%%%%%%%%%%%%%%%%%%%%%%%%%%%%
		\begin{scope}[xshift=6cm]
		\node[name=Z, shape=swig vsplit]{ %Construct a vertically split "node"
			\nodepart[est]{left}{$Z$} % left half
			\nodepart[red]{right}{$z$}
			};
		\node[name=D, shape=swig vsplit, right=0.65cm of Z]{  %Construct a vertically split "node"
			\nodepart[est]{left}{$D({\color{red}{z}})$}  % left half
			\nodepart[red]{right}{$d$} % right half
			}; 
		\node[est, name=Y,shape=ellipse,style={draw}, right =0.65cm of D,inner sep=2pt]{$Y({\color{red}{d}})$};
		\draw[pil,->] (Z) to (D);
		\node[est, above =1cm of D] (X) {$X$};
		\path[pil] (Z) edgenode {} (D);
		\path[pil] (D) edgenode {} (Y);
		\draw[pil, ->] (X)  to[bend right]  (Z.140);
		\draw[pil, ->] (X)  to  (D.110); %D.110 is angle 110
		\path[pil] (X) edgenode {} (Y);
		\node[shade, below = of D] (U) {$U$};
		\path[pil] (U) edgenode {} (D.250);% D.250 is angle 250
		\path[pil] (U) edgenode {} (Y);
		\node[name=la, below=2cm of D]{(b). A Single World Intervention Graph.};
		\end{scope}
		\end{tikzpicture}	
	\caption{Illustration of an instrumental variable model using a causal graph.  Variables $X,Z,D,Y$ are observed; $U$ is unobserved. 
	The left panel gives a causal Directed Acyclic Graph \citep{pearl2009causality}, 
	and the right panel gives a Single World Intervention Graph \citep{richardson2013single}}
	\label{DAG:iv_model}
\end{figure}

Under the principal stratum framework \citep{frangakis2002principal},  the population can be divided into four strata based on values of $(D(1), D(0))$ as in Table \ref{table:defG}. We use $t_D$ to denote principal stratum defined by values of $(D(1), D(0))$.
\begin{table}[!ht]
	\centering
	\caption{Principal stratum $t_D$ based on $(D(1),D(0))$  }\label{table:defG}
	\begin{tabular}{cc|cc}
		\toprule
		$D(1)$ & $D(0)$ & Principal stratum & Abbreviation \\
		\midrule
		1 & 1 &  Always taker & AT \\
		1 & 0 &  Complier  & CO \\
		0 & 1 &  Defier & DE \\
		0 & 0 &  Never taker & NT \\
		\bottomrule
	\end{tabular}
\end{table}
We are interested in estimating the conditional treatment effects in the complier stratum on the additive and multiplicative scales defined as
\begin{flalign*}
\mathrm{LATE}(X) &= E[Y(1)-Y(0)\mid D(1)> D(0),X]; \\
\mathrm{MLATE}(X) &= E[Y(1)\mid D(1)> D(0),X] / E[Y(0)\mid D(1)> D(0),X].
\end{flalign*} 
For $\mathrm{MLATE}(X)$ to be well-defined, we also assume that $E[Y(0)\mid D(1)> D(0),X] \neq 0$ almost surely.  By definition, the parameter space of $\mathrm{LATE}(X)$ and $\mathrm{MLATE}(X)$ are constrained: $\mathrm{LATE}(X) \in [-1,1]$ while $\mathrm{MLATE}(X) \in [0,+\infty).$

\citet[][Lemma 2.1]{abadie2002bootstrap} shows that under assumptions A1-A5, the local average treatment effects are identifiable as
\begin{flalign}
	\mathrm{LATE}(X) &= \delta^L(X) \equiv \dfrac{E(Y\mid Z=1,X) - E(Y\mid Z=0,X)}{E(D\mid Z=1,X) - E(D\mid Z=0,X)}; \label{eqn:abadie:late} \\
	\mathrm{MLATE}(X) &= \delta^M(X) \equiv -\dfrac{E(YD\mid Z=1,X) - E(YD\mid Z=0,X)}{E\{Y(1-D)\mid Z=1,X\} - E\{Y(1-D)\mid Z=0,X\}}.  \label{eqn:abadie:mlate}
\end{flalign}
%The formulas \eqref{eqn:abadie} also follow from equations \eqref{eqn:whole_2} and \eqref{eqn:whole_3} in Section \ref{sec:parameterization}.

Given \eqref{eqn:abadie:late}, it might be tempting to estimate $\delta^L(X)$ and hence $\mathrm{LATE}(X)$ with a plug-in estimator by first estimating the four curves $E(Y\mid Z=z,X)$ and $E(D\mid Z=z,X), z=0,1$ separately, as proposed by \cite{frolich2007nonparametric}. However, even though one may choose suitable models so that estimates for the four curves above lie in the unit interval, there is no guarantee that the plug-in estimator is between -1 and 1. The same problem arises when applying  \cite{tan2006regression}'s approach that imposes parametric models on the conditional means $E(Y\mid D=d, Z=z,X)$ and $E(D\mid Z=z,X), d,z=0,1$.  Similar problems arise with plug-in estimators for $\delta^M(X)$.

To avoid these problems, \cite{abadie2003semiparametric} specifies a parametric model, such as the logistic model, for the so-called local average response function $E\{Y(d)\mid D(1)>D(0),X\}.$ The model parameters are then estimated by a weighted estimating equation. The parameters of  logistic models, however, do not directly encode the dependence of local average treatment effects on baseline covariates, so they do not offer direct insights into which baseline variables  modify the local average treatment effects.   Moreover, the validity of their approach hinges on correct specification of the instrumental density $P(Z=1\mid X)$.  Instead,  \cite{okui2012doubly} and \cite{ogburn2015doubly} propose doubly robust estimators based on  direct parameterization of the target functional  $\delta^L(X)$. Given a correct model $\delta^L(X;\alpha)$,  their estimators are consistent and asymptotically normal for the parameter of interest $\alpha$  if either the instrumental density model $P(Z=1\mid X;\gamma)$ or another nuisance  model  $E(Y-D \times  \delta^L(X)\mid X; \beta)$ is correctly specified. However,  the nuisance model $E(Y-D \times  \delta^L(X)\mid X; \beta)$ is variation dependent of the target model $\delta^L(X; \alpha)$. In this case, the double robustness property of \cite{okui2012doubly} and \cite{ogburn2015doubly}'s estimators is not practically meaningful since with continuous covariates, often it is not possible for  $\delta^L(X; \alpha)$ and $E(Y-D \times  \delta^L(X)\mid X; \beta)$ to be correct simultaneously. Similar discussions apply to the target functional $\delta^M(X).$

\hlc{In related work, \cite{wang2018bounded} studied the problem of estimating a closely related functional, $E_X\{ \delta^L (X)\}$, which may be interpreted as the average treatment effect under a certain set of identification assumptions.} As an intermediate step, \citet[][\S 4.1]{wang2018bounded} propose  alternative nuisance models that are variation independent of $\delta^L(X; \alpha),$ including a model for  $\delta^D(X) \equiv E(D\mid Z=1,X) - E(D\mid Z=0,X)$.  
The key observation made by these authors is that as long as the models for $\delta^L(X)$ and $\delta^D(X)$ both lie in their parameter space, which is $[-1,1]$ for $\LATE(X)$ and $[0,1]$ for $\delta^D(X)$, then 
$$
		E(Y\mid Z=1,X) - E(Y\mid Z=0,X) = \LATE(X) \times \delta^D(X)
$$
also lies in its parameter space $[-1,1].$ Following this, they derive a maximum likelihood estimator \citep[][\S 4.1]{wang2018bounded} and a truly doubly robust estimator  for $\delta^L(X)$  \citep[equation (14)]{wang2018bounded}. 
These approaches, however, cannot be adapted to estimate $\delta^M(X)$ or the multiplicative local average treatment effect.   Furthermore, as we explain later in Remark \ref{remark:optimal}, even for the additive local average treatment effect, in general their doubly robust estimator fails to achieve the semiparametric efficiency bound.

%obtain truly doubly robust estimators for $MLATE(X)$ as $MLATE(X) \in [0,+\infty)$, 
%$E\{Y(1-D)\mid Z=1,X\} - E\{Y(1-D)\mid Z=0,X\} \in [-1,1]$ while $E\{YD\mid Z=1,X\} - E\{YD\mid Z=0,X\} \in [-1,1].$  Furthermore, $MLATE(X)$ is variation dependent of $E\{Y(1-D)\mid Z=1,X\} - E\{Y(1-D)\mid Z=0,X\}.$ 

\section{A novel parameterization}
\label{sec:parameterization}

In this section we describe a novel parameterization of the {observed data likelihood} involving $\LATE(X; \alpha)$ or $\MLATE(X; \alpha).$  
Specifically, our goal is to find nuisance models so that (I) they are variation independent of  each other; (II) they are variation independent of  $\LATE(X;\alpha)$ and $\MLATE(X; \alpha)$; (III) there exists a bijection between the observed data likelihood on $P(D=d,Y=y\mid Z=z,X)$ and the combination of  target and nuisance models. Note the remaining parts of the likelihood on $P(Z=z, X)$ do not show up in the identification formula \eqref{eqn:abadie:late} or \eqref{eqn:abadie:mlate}. Thus they contain no information about the parameters of interest and need not be modeled. 

Let $p_x(d,y\mid z) = P(D=d, Y=y\mid Z=z, X=x).$ For any $x$, the parameter space of the observed data likelihood on $p_x(d,y\mid z)$ is a six-dimensional space in $[0,1]^8$ \citep{richardson2011transparent}:
\begin{align*}
\MoveEqLeft{\Delta =\left\{p_x(d,y\mid z) \geq 0:  \sum\limits_{d,y} p_x(d,y\mid z) = 1,\right.}\\[-8pt]
&\quad\quad\quad\quad\quad
\quad\left.
\vphantom{\sum\limits_{d,y}}%to get parentheses sized
p_x(1,y\mid 1) \geq p_x(1,y\mid 0),\; p_x(0,y\mid 1) \leq p_x(0,y\mid 0),\; y=0,1.
%\\
%& \left. p_x(d,y\mid 1)+ p_x(d,1-y\mid 0)\leq 1, d=0,1, y=0.1.
\right\} \numberthis\label{eqn:delta}
\end{align*}
Parameterization of $p_x(d,y\mid z)$ is a difficult problem since as shown by  \eqref{eqn:delta},  the likelihood components $p_x(d,y\mid z), d,y,z=0,1$ are not variation independent of each other. 

To make progress, instead of modeling the observed likelihood components directly, we seek to model components of the potential outcome likelihood $(D(1), D(0), Y(1), Y(0))$ conditional on $X$. Specifically, we will consider $p(AT;X) \equiv P(t_D=AT\mid X), p(NT;X), p(CO;X)$ and $p(Y(1)\mid AT;X) \equiv P(Y(1)=1\mid t_D=AT, X), p(Y(0)\mid NT;X), p(Y(1)\mid CO;X)$ and $p(Y(0)\mid CO;X)$. The remaining parts of the potential outcome likelihood, such as $p(Y(1)\mid NT;X)$ are not modeled as they are not related to the observed data likelihood, and hence contain no information for the parameters of interest. These components, however, are still variation dependent since 
\begin{equation}
\label{eqn:constraint}
p(AT;X) + p(NT;X) + p(CO;X) = 1.
\end{equation}	
Moreover, they do not contain our target function $\LATE(X)$ or $\MLATE(X),$ which we denote as $\theta(X)$.

Theorem \ref{thm:para} presents an alternative parameterization that satisfies goals (I)--(III).  To avoid the constraint \eqref{eqn:constraint}, we follow \cite{wang2017identification} to re-parameterize $p(AT;X), p(NT;X), p(CO;X)$. To re-parameterize $p(Y(1)\mid CO;X)$ and $p(Y(0)\mid CO;X)$ so that the new parameterization includes $\theta(X)$, we  follow  \cite{richardson2017modeling} to model an odds product function in the complier stratum. 
%Variation independence of the nuisance functions and the target function follows directly from the fact that their joint range is a Cartesian product of their marginal ranges.   
 The proof of Theorem \ref{thm:para} is left to the Supplementary Material. 

\begin{theorem}
	\label{thm:para}
 Let
 $\m$ denote the 6-dimensional models consisting of the target model $\theta(X; \alpha)$ and models on the following nuisance functions:
	\begin{flalign*}
		\phi_1(X) &\equiv P(t_D = CO\mid X) &&= P(D\!=\!1\,|\, Z\!=\!1, X) - P(D\!=\!1\,|\, Z\!=\!0,X); \\
		\phi_2(X) & \equiv P(t_D = AT\mid t_D\in \{AT, NT\}, X) &&= \dfrac{P(D=1\mid Z=0,X)}{P(D\!=\!1\,|\, Z\!=\!0, X) + P(D\!=\!0\,|\, Z\!=\!1, X)}; \\
	\phi_3(X) &\equiv  P(Y=1\mid t_D = NT, X) &&= P(Y=1\mid D=0,Z=1,X);\\
	\phi_4(X) &\equiv   P(Y=1\mid t_D = AT, X) &&= P(Y=1\mid D=1, Z=0, X);
	\end{flalign*}
	\begin{flalign*}
 OP^{CO}(X) &\equiv  \dfrac{E\{Y(1)\mid t_D = CO,X\}   E\{Y(0)\mid t_D = CO,X\} }{[1-E\{Y(1)\mid t_D = CO,X\}] [1-E\{Y(0)\mid t_D = CO,X\}] },
	\end{flalign*}
where $OP^{CO}$ denotes odds product in the complier stratum.

Under assumptions A1 - A5, for any realization of $X$, the map given by 
\begin{flalign}
\label{eqn:map}
	(P(D=d, Y=y\mid Z=z,X), d,y,z\in\{0,1\}) \rightarrow	(\theta(X), &\phi_1(X), \phi_2(X), \phi_3(X), \phi_4(X), OP^{CO}(X) ) 
\end{flalign}
is well-defined and is a smooth bijection from $\Delta$ to $\mathcal{D} \times [0,1]^4 \times [0,\infty)$, where $\mathcal{D} = [-1,1]$ if $\theta(X) = \LATE(X)$ and $\mathcal{D}=[0,\infty)$ if $\theta(X) = \MLATE(X).$ Furthermore, the models in $\m$ are variation independent of each other. 
\end{theorem}

Suppose models for $\theta(X), \phi_1(X), \ldots, \phi_4(X), OP^{CO}(X)$ are all specified up to a finite dimensional parameter, then these parameters, and in particular the local average treatment effects may be estimated directly via unconstrained maximum likelihood based on the diffeomorphism  \eqref{eqn:map}. Likelihood-based confidence intervals  can then be obtained in standard fashion.

\begin{remark}
\hlc{Since the constituent models, $\theta(X), \phi_1(X), \ldots, \phi_4(X), OP^{CO}(X)$, are variation independent, the modeler is free to pick any function of $X$ with the given range. For example, to mitigate model mis-specification, one may assume flexible machine learning models on these functions of $X$. In this case, one can similarly fit these flexible models based on the implied models on the likelihood $P(D=d, Y=y\mid Z=z, X).$}
\end{remark}

\section{Doubly Robust Estimation}

In this section we apply our parameterization in Theorem \ref{thm:para} to construct truly  doubly robust estimators that are asymptotically linear for estimating the local average treatment effects if either the nuisance models $\phi_1(X; \beta_1), \ldots, \phi_4(X; \beta_4), OP^{CO}(X; \eta)$ or the instrumental density model $P(Z=1\mid X; \gamma)$ is correct, given that the causal model $\theta(X; \alpha)$ is correctly specified.  These estimators are called truly doubly robust because, as shown in Theorem \ref{thm:para}, the nuisance models and causal model are variation independent  and hence congenial to each other.

Let $\hat{\alpha}, \hat{\beta}_{1}, \ldots, \hat{\beta}_{4}, \hat{\eta}$ and  $\hat{\gamma}$  be the maximum likelihood estimators of $\alpha, \beta_1, \ldots, \beta_4, \eta, \gamma$. Also let 
\[
H(Y,D,X;\alpha)=
\begin{cases}
Y - D \theta(X;\alpha)& \theta(X) = \LATE(X);\\
Y \theta(X;\alpha)^{-D}	& \theta(X) = \MLATE(X).
\end{cases}
\]
%One may show that the causal model can be expressed as the conditional moment model \citep{robins1994correcting}:
%$$
%		E\{H(Y,D,X;\alpha)\mid Z,X\} = E\{H(Y,D,X;\alpha) \mid X \}.
%$$
We have the following theorem.

\begin{theorem}
	\label{thm:doublyrobust}
	Let $\hat{\alpha}_{dr}$ solve the following estimating equation:
	\begin{equation}
	\label{eqn:ee}
	\mathbb{P}_n \omega(X) \dfrac{2Z-1}{f(Z\mid X; \hat{\gamma})} \left[H(Y,D,X;\alpha) - \hat{E}\left\{ H(Y,D,X;\alpha) \mid X\right\} \right] = 0,
	\end{equation}
	where
	\[
		f(Z\mid X; \hat{\gamma})=\{P(Z=1\mid X; \hat{\gamma})\}^Z \{1-P(Z=1\mid X; \hat{\gamma})\}^{1-Z};
	\]
	\[
	\hat{E}\left\{ H(Y,D,X;\alpha) \mid X\right\}=
	\begin{cases}
	 \hat{f}_0 \hat{\phi}_1 + (1-\hat{\phi}_1) (1-\hat{\phi}_2) \hat{\phi}_3 +  (1-\hat{\phi}_1) \hat{\phi}_2 \hat{\phi}_4 -\theta(1-\hat{\phi}_1) \hat{\phi}_2 & \theta(X) = \LATE(X);\\
	 \hat{f}_0 \hat{\phi}_1 + (1-\hat{\phi}_1) \hat{\phi}_2 \hat{\phi}_4 {\theta}^{-1} +  (1-\hat{\phi}_1) (1-\hat{\phi}_2) \hat{\phi}_3 & \theta(X) = \MLATE(X);
	\end{cases}
	\]
	with 
	\[
	\hat{f}_0=
	\begin{cases}
	\dfrac{1}{2(\widehat{OP}-1)} \left\{ 
	\widehat{OP}(2-\theta)+\theta-  	\sqrt{{\theta}^2(\widehat{OP}-1)^2 + 4\widehat{OP}}\right\}& \theta(X) = \LATE(X);\\
	\dfrac{1}{2{\theta}(1-\widehat{OP})} \left\{ 
	-({\theta}+1)\widehat{OP}+  	\sqrt{\widehat{OP}^2({\theta}-1)^2 + 4{\theta}\widehat{OP}}\right\} & \theta(X) = \MLATE(X);
	\end{cases}
	\]
	and 
	\[
		\theta =  \theta(X;\alpha);  \quad \hat{\phi}_i = \phi(X;\hat{\beta}_i), i=1,\ldots,4; \quad 	\widehat{OP} =  OP^{CO}(X; \hat{\eta});
	\]
$\mathbb{P}_n$ denotes the empirical mean operator and $\omega(X)$ is an arbitrary measurable function of $X$.  Then under a correct model for $\theta(X;\alpha)$ and regularity conditions, $\hat{\alpha}_{dr}$ is consistent and asymptotically normally distributed provided that at least one of the models for ${E}\left\{ H(Y,D,X;\alpha) \mid X\right\}$ or $f(Z\mid X;\gamma)$ is correctly specified. The optimal choice of $\omega(X)$ that minimizes the asymptotic variance of $\hat{\alpha}_{dr}$ is given in the Supplementary Material.
	\end{theorem}
	
% The proof of Theorem	\ref{thm:doublyrobust} is given in the Supplementary Material. It is worth mentioning that  the arguments of square roots in $\hat{f}_0$ are always positive.

Theorem \ref{thm:doublyrobust} is a special case of the  doubly robust g-estimation theory developed by \citet[][\S 3.2, \S 3.3]{ogburn2015doubly}. For completeness, we provide the proof in the Supplementary Material. One can also easily verify that the arguments in the square roots in $\hat{f}_0$ are always non-negative, provided that the estimates of $\theta(X)$ and $OP^{CO}(X)$ stay within their respective domain.
Statistical inference may be based on standard M-estimation theory. Alternatively, in the simulations and real data analysis, we use nonparametric bootstrap. 

\begin{remark}
	\label{remark:optimal}
When $\theta(X) = \LATE(X),$	\cite{wang2018bounded}'s approach parameterizes the marginal distributions $P(Y=1\mid Z=z, X)$ and $P(D=1\mid Z=z,X)$, whereas our approach parameterizes the joint distribution $P(D=d, Y=y\mid Z=z, X).$  On the other hand, 	in addition to the marginal distributions, the optimal choice of $\omega(X)$ also depends on  $P(DY=1\mid Z=z, X)$. Hence it  can be calculated based on our parameterization but not  \cite{wang2018bounded}'s. 
\end{remark}

\begin{remark}
\label{remark:simple}
   \hlc{ Prompted by the comment of a referee,} we notice that if $\theta(X) = \delta^L(X),$ then $E\{H(Y,D,X;\alpha)\mid X\} = E(Y\mid X) - \theta(X;\alpha)E(D\mid X) $. Hence a simple way to estimate $\alpha$ based on \eqref{eqn:ee} is to first obtain estimates of $E(Y\mid X),E(D\mid X)$ and $f(Z\mid X)$ and then plug these estimates into \eqref{eqn:ee} to estimate $\alpha$. Moreover, in the Supplementary Material, we show that $E(Y\mid X),E(D\mid X),$ $f(Z\mid X)$ and $\delta^L(X)$ are variation independent in the interior of their domains; a similar phenomenon was previously observed by \citet[][\S 3.1]{wang2017congenial}. Similarly, if $\theta(X) = \delta^M(X),$ then $E\{H(Y,D,X;\alpha)\mid X\} = {P}(D=1\mid X) {P}(Y=1 \mid D=1, X) \theta(X;\alpha)^{-1} + {P}(D=0\mid X) {P}(Y=1\mid D=0,X).$ Hence a simple way to estimate $\alpha$ based on \eqref{eqn:ee} is to first obtain estimates of $E(Y\mid X), E(Y\mid D,X)$ and $f(Z\mid X)$, and then plug these estimates into \eqref{eqn:ee} to estimate $\alpha$.  $E(Y\mid X), E(Y\mid D,X), f(Z\mid X)$ and $\delta^M(X)$ are also variation independent in the interior of their domains. Consequently,
   since all the nuisance models can, logically, be correctly specified,
   these simple estimators are truly doubly robust assuming that the true parameter values are away from the boundary. However,
   similar to the estimator of \citet{wang2018bounded},
   they cannot be used to estimate the optimal $\omega(X)$. Furthermore, these simple parameterizations do not lead to likelihood-based inference.
\end{remark}

%\change{Need a formal theorem stating double robustness and efficiency?}

%Theorem \ref{thm:dr_estiamtion} summarizes the key properties of $\hat{\alpha}_{dr}$.
%
%\begin{theorem}
%	\label{thm:dr_estiamtion}
%\end{theorem}

	\section{Simulation studies}
\label{sec:simulations}
In this section, we evaluate the finite sample performance of various estimators discussed in this paper. We generate data from the following models:
\begin{flalign*}
\LATE(X) &= {\rm tanh}(\alpha^\top X); \hspace{6.5em} \MLATE(X)=\exp (\alpha^\top X);\\
\phi_i(X)&={\rm expit}(\beta_i^\top X), i=1,\ldots,4; \quad OP^{CO}(X)=\exp (\eta^\top X);\\
P(Z=1\mid X) &= {\rm expit}(\gamma^\top X),
\end{flalign*}
where the covariates $X$ include an intercept and a random variable generated from ${\rm Unif}(-1,1)$; $\alpha=(0,-1)^\top$, $\beta_i=(-0.4,0.8)^\top$, $i=1,\ldots,4$, $\eta=(-0.4,1)^\top$ and $\gamma=(0.1,-1)^\top$. Under this setting, the strength of instrumental variable, defined as $\Delta^D=E\{E(D\mid Z=1,X)-E(D\mid Z=0,X)\}$, is $0.406$. The sample size is $1000$.

We also consider scenarios in which the nuisance models are mis-specified. In these scenarios, the analyst is given covariates $X^\dagger$ including an intercept and an irrelevant covariate generated from an independent ${\rm Unif}(-1,1)$, and  covariates $X^\prime$ including 
\[
(\underbrace{1,\ldots,1}_{0.5n},\underbrace{0\ldots,0}_{0.5n})^\top
 \quad \text{and}\quad (\underbrace{0,\ldots,0}_{0.1n},\underbrace{1,\ldots,1}_{0.9n})^\top.
 \]
%\tblue{Is there a better way to write this?} 
Instead of formulating a model conditioning on $X$, the analyst  fits the model $P(Z=1\mid X^\dagger;\gamma)$ and/or $\phi_i(X^\prime;\beta_i)$, $i=1,\ldots,4$, ${ OP}^{CO}(X^\prime;\eta)$. The analyst still uses the correct functional form in these models. The target model $\theta(X;\alpha)$ is always correctly specified. Figure S1 in the Supplementary Material visualizes the degree of model mis-specification by showing the data points generated under the true models and mis-specified models from one randomly selected Monte Carlo run.

We consider the performance of following estimators:
 \begin{itemize}
 \item[{\tt {mle}}:] the proposed maximum likelihood estimator;
 \item[{\tt {drw}}:] the proposed doubly robust estimator with the optimal weighting function;
\item[{\tt {dru}}:] the proposed doubly robust estimator with the identity weighting function;
\item[{\tt {reg.ogburn}}:] \citet[\S 3.1]{ogburn2015doubly}'s outcome regression estimator;
\item[{\tt {drw.ogburn}}:]  \citet[\S 3.3]{ogburn2015doubly}'s doubly robust estimator with the optimal weighting function;
\item[{\tt {dru.ogburn}}:]  \citet[\S 3.2]{ogburn2015doubly}'s doubly robust estimator with the identity weighting function;
\item[{\tt {mle.wang}}:]  \citet[\S 4.1]{wang2018bounded}'s maximum likelihood estimator;
\item[{\tt {dru.wang}}:]  \citet[\S 4.4]{wang2018bounded}'s doubly robust estimator with the identity weighting function;
\item[{\tt {dru.simple}}:]  The doubly robust  estimator described in Remark \ref{remark:simple};
\item[{\tt {ls.abadie}}:] \citet[\S 4.2.1]{abadie2003semiparametric}'s least squares estimator;
 \item[{\tt {mle.crude}}:] \citet[\S 2]{richardson2017modeling}'s maximum likelihood estimator of the crude association on the additive/multiplicative scale.
  \end{itemize}
For models other than the proposed ones, we provide   details of model specifications in the Supplementary Material.
 
% \noindent${\tt mle.crude}$: \cite{richardson2017modeling}'s maximum likelihood estimator that only accounts for observed confounding \tlube{and estimates the average treatment effect;}\\
% ${\tt mle}$: the proposed maximum likelihood estimator;\\
% ${\tt drw}$: the proposed doubly robust estimator with the optimal weighting function;\\
% ${\tt dru}$: the proposed doubly robust estimator with the identity weighting function;\\
% ${\tt reg.ogburn}$: \cite{ogburn2015doubly}'s outcome regression estimator;\\
% ${\tt drw.ogburn}$: \cite{ogburn2015doubly}'s doubly robust estimator with the optimal weighting function;\\
% ${\tt dru.ogburn}$: \cite{ogburn2015doubly}'s doubly robust estimator with the identity weighting function;\\
% ${\tt mle.wang}$: \cite{wang2018bounded}'s regression-based estimator;\\
% ${\tt dru.wang}$: \cite{wang2018bounded}'s doubly robust estimator with the identity weighting function;\\ 
% ${\tt dru.simple}$: a simple  estimator proposed by a referee based on estimating equation \eqref{eqn:ee}  using the identity weighting function and parameterizing $E(D|X)$ and $E(Y|D,X)$; see the Supplementary Material for a detailed description; 
% ${\tt ls.abadie}$: \cite{abadie2003semiparametric}'s least squares estimator.

Throughout our simulations,  we assume the model of interest $\theta(X)$ is always corrected specified. We consider the following four scenarios for the nuisance models:
 \begin{itemize}
 \item[{\tt {bth}}:] $X$ is used in all nuisance models;
 \item[{\tt {psc}}:] $X$ is used in the instrumental density model, but $X^{\prime}$ is used in other nuisance models;
 \item[{\tt {opc}}:] $X^{\dagger}$ is used in the instrumental density model, but $X$ is used in other nuisance models;
\item[{\tt {bad}}:]$X^{\dagger}$ is used  in the instrumental density model, and $X^\prime$ is used in other nuisance  models.
  \end{itemize}
% \noindent {\tt bth}: $X$ is used in all models;\\
% {\tt psc}: $X$ is used in the instrumental density model, but $X^\prime$ is used in other models except $\theta(X)$;\\
% {\tt opc}: $X^{\dagger}$ is used in the instrumental density model, but $X$ is used in other models;\\
% {\tt bad}: $X^{\dagger}$ is used  in the instrumental density model, and $X^\prime$ is used in other models except $\theta(X)$.
As \cite{abadie2003semiparametric}'s method does not directly  specify a model for $\theta(X)$, we consider the following two scenarios for their method:
 \begin{itemize}
 \item[{\tt {bth}}:] $X$ is used in all models;
\item[{\tt {bad}}:] $X^{\dagger}$ is used in the instrumental density model, but $X$ is used in the model for $E\{Y\mid X,D,D(1)>D(0)\}$.
  \end{itemize}
 The implied model for $\theta(X)$ remains correct in either of these two scenarios.
  
% \noindent {\tt bth}: $X$ is used in all models;\\
% {\tt bad}: $X^{\dagger}$ is used in the instrumental density model, but $X$ is used in the model for $E\{Y\mid X,D,D(1)>D(0)\}$.

% Although some working models are wrong, they are treated as true models and fitted based on the original methods. 
% Note that because the instrumental density  model is not used in {\tt mle}, results for {\tt mle.opc} are identical to {\tt mle.bth}, and results for {\tt mle.psc} are identical to {\tt mle.bad}. The conclusions are the same for {\tt reg.ogburn} and {\tt mle.wang}. 
Table \ref{tab:biassd} presents selected results of  the bias and the Monte-Carlo standard error for various estimators based on $1000$ Monte-Carlo runs. In Section 5.3 of the Supplementary Material, we present the complete set of results  in Table S2, and bias as a percentage of the estimator's standard deviation in Table S3. The estimator {\tt mle.crude} has large bias, indicating that the effect of unmeasured confounding is non-negligible.
As expected, the proposed estimators have small bias relative to standard error in all scenarios except for {\tt mle.bad} and {\tt drw.bad}. As expected, when all nuisance
models are correctly specified, the  proposed maximum likelihood estimator has smaller or comparable
standard error to the optimally-weighted doubly robust estimator {\tt drw.bth}. 
The performance of {\tt dru.ogburn.bth}, {\tt dru.wang.bth} and {\tt dru.simple.bth} are all similar to that of {\tt dru.bth}; all these four methods are less efficient than {\tt drw.bth}. This suggests that  under our simulation settings, the optimal weighting function leads to important  efficiency gain. Although {\tt drw.ogburn.bth} is constructed  based on the same optimally weighted estimating equation as {\tt drw.bth},  mis-specification of variation dependent models leads to a biased estimate of the optimal weight function. As a result, in some cases,  
it is even less efficient than {\tt dru.ogburn.bth}.

\begin{table}[!htbp]
	\centering
	\caption{The biases and standard errors of the estimated biases in the Monte-Carlo study of various estimators in the selected scenarios. The true value for $\alpha_0$ and $\alpha_1$ is 0 and -1, respectively. The sample size is  1000}
	\bigskip
	\small
    \label{tab:biassd}
	\begin{tabular}{rccccccc}
	\toprule
&       \multicolumn{2}{c}	{$\theta(X) = \LATE(X)$} &  \multicolumn{2}{c}	{ $\theta(X) = \MLATE(X)$} \\
\cmidrule(r){2-3} \cmidrule(l){4-5}
& \multicolumn{1}{c}{$\alpha_0$} & \multicolumn{1}{c}{$\alpha_1$} &   \multicolumn{1}{c}{$\alpha_0 $} & \multicolumn{1}{c}{$\alpha_1 $} \\
\midrule
\multicolumn{2}{l}{Bias $\times 100$ (SE $\times 100$) \quad \quad}	  & & \\ [5pt]
	{\quad \quad}	mle.bth & 0.28(0.35) & -3.5(0.78) & -0.092(0.71) & -3.0(1.2) \\ 
mle.bad & -20(0.42) & -15(0.80) & -48(1.2) & -18(2.1) \\ [3pt]
drw.bth & 0.55(0.36) & -4.1(0.82) & 0.54(0.77) & -5.6(1.5) \\ 
drw.psc & 0.060(0.38) & -5.9(1.0) & -0.38(1.2) & -12(2.7) \\ 
drw.opc & 0.55(0.36) & -3.9(0.79) & 0.49(0.75) & -5.3(1.4) \\ 
drw.bad & -10(0.40) & -9.6(1.1) & -28(1.4) & 25(3.3)  \\[3pt] 
dru.bth & 1.3(0.44) & -5.8(1.0) & 1.8(0.84) & -8.1(1.7) \\  [5pt]
reg.ogburn.bth & -5.7(1.6) & -2.9(3.1) & 7.8(2.0) & -1.1(2.2) \\ 
reg.ogburn.bad & -9.0(0.25) & 100(0.23) & 140(5.6) & 93(3.6) \\ [3pt] 
drw.ogburn.bth & 0.10(0.46) & -4.2(0.99) & 3.2(1.4) & -13(2.5) \\ 
 [3pt]
dru.ogburn.bth & 1.3(0.45) & -5.8(1.1) & 1.9(0.85) & -8.2(1.7) \\   [5pt]
dru.wang.bth & 1.3(0.45) & -5.8(1.0) & $-$ & $-$ \\    [5pt]
dru.simple.bth & 1.3(0.45) & -5.8(1.0) & 1.8(0.84) & -8.0(1.7) \\ 
dru.simple.psc & 1.2(0.44) & -6.2(1.0) &1.9(0.84) & -8.8(1.7) \\ 
dru.simple.opc & 4.5(0.49) & -17(1.2) & -0.15(0.68) & 11(1.2) \\ 
dru.simple.bad & -16(0.48) & -17(1.3) & -34(0.70) & 18(1.5) \\ [5pt]
ls.abadie.bth & -0.19(0.37) & -4.1(0.93) & 0.42(0.79) & -11(1.6) \\ 
ls.abadie.bad & -23(0.88) & 22(1.2) & -32(1.9) & 7.7(3.6) \\  [5pt]
mle.crude & -2.8(0.10) & 60(0.19) & 0.36(0.25) & 51(0.42) \\ 
			\bottomrule 
	\end{tabular}
\end{table}

Table \ref{tab:ci} reports selected coverage probabilities  of 95\% confidence intervals obtained from quantile bootstrap based on 500 bootstrap samples, with the complete set of results presented in Table S4 in the Supplementary Material. The proposed estimators has coverage close to the nominal level  except that for  {\tt mle.bad} and {\tt drw.bad}. Inference results produced by {\tt reg.ogburn.bth} and {\tt drw.ogburn.bth} tend to be overly conservative, \hlc{possibly due to mis-specification of variation dependent models.}

\begin{table}
	\centering
	\caption{The coverage probabilities of confidence intervals obtained from $500$ bootstrap samples in selected scenarios. The true values for $\alpha_0$ and $\alpha_1$ are 0 and -1, respectively. The sample size is  1000}
	\bigskip
	\small
    \label{tab:ci}
	\begin{tabular}{rcccccc}
	\toprule
&       \multicolumn{2}{c}	{$\theta(X) = \LATE(X)$} &  \multicolumn{2}{c}	{ $\theta(X) = \MLATE(X)$} \\
\cmidrule(r){2-3} \cmidrule(l){4-5}
& \multicolumn{1}{c}{$\alpha_0$} & \multicolumn{1}{c}{$\alpha_1$} &   \multicolumn{1}{c}{$\alpha_0 $} & \multicolumn{1}{c}{$\alpha_1 $} \\
\midrule
\multicolumn{2}{l}{Coverage probability $\times 100$ \quad \quad}	  & & \\ [5pt]
	{\quad \quad}	mle.bth & 95.6 & 95.8 & 95.4 & 96.4 \\ 
mle.bad & 65.4 & 91.0 & 46.3 & 94.6 \\ [3pt] 
drw.bth & 94.7 & 95.2 & 96.9 & 95.9 \\ 
drw.psc & 95.4 & 95.5 & 97.6 & 97.4 \\ 
drw.opc & 95.0 & 95.6 & 96.1 & 96.1 \\ 
drw.bad & 87.0 & 95.3 & 91.8 & 96.8 \\  [3pt] 
dru.bth & 94.5 & 94.6 & 96.3 & 96.9 \\  [5pt] 
reg.ogburn.bth & 98.0 & 99.9 & 99.6 & 100.0 \\ 
reg.ogburn.bad & 75.6 & 0.1 & 99.9& 86.1 \\  [3pt]
drw.ogburn.bth & 97.0 & 98.1 & 98.5 & 98.4 \\ [3pt]
dru.ogburn.bth & 94.5 & 95.0 & 96.2 & 97.2 \\   [5pt]
dru.wang.bth & 94.4 & 94.7 &  $-$ &  $-$  \\   [5pt]
dru.simple.bth & 94.3& 94.8 &  96.1& 96.5 \\ 
dru.simple.psc & 94.7 & 94.6  & 96.1 & 96.8\\ 
dru.simple.opc & 93.6 & 92.6 & 96.2 & 94.9 \\  
dru.simple.bad & 76.3 & 94.4 & 69.3 & 93.8 \\  [5pt]
ls.abadie.bth & 94.8 & 95.5& 96.4 & 95.6 \\ 
ls.abadie.bad & 87.3 & 94.5 & 93.1 & 95.7 \\[5pt]
mle.crude & 84.3 & 0.0& 94.0 & 6.3 \\ 
			\bottomrule 
	\end{tabular}
\end{table}

\section{Application to 401(k) data}

We apply the proposed procedures to evaluate the effect of 401(k) retirement plan on savings, which has become the most popular employer-sponsored retirement plan in the United States. 
 Economists have long been interested in whether 401(k) contributions represent additional savings or simply replace other retirement plans, such as Individual Retirement Accounts.  
  To account for unobserved confounders such as the underlying preference for savings,  \cite{abadie2003semiparametric} chooses 401(k) eligibility as an instrument.  Since eligibility is determined by employers, individual preferences for savings may play a minor role in the determination of eligibility after controlling for observed covariates including  family income, age,  marital status and family size. Furthermore, it is plausible that 401(k) eligibility  has an impact on participation in Individual Retirement Accounts only through participation in 401(k) plans. The monotonicity and instrumental variable relevance assumptions hold trivially as only eligible individuals may choose to participate in 401(k) plans. 
%   In \cite{abadie2003semiparametric},  they proved that the instrument $Z$,  401(k) eligibility,  was associated with family income, age and marital status. Eligibles have higher average income and they are more likely to be married. So the assumption of the direction from baseline covariates $X$ to instrument $Z$ is likely to hold.
  
  In our analysis, we use the data set prepared for \cite{abadie2003semiparametric}, which contains 9275 individuals from the Survey of Income and Program Participation  of 1991.  
  The study participants were between 25 and 64 years old, had an annual income between \$10,000 and \$200,000 and  a family size ranging from 1 to 13;   62.9\% of them were married.
  The assumptions A1, A2 of the instrumental variable model imply restrictions on the observed data law \citep{pearl:testability:95}.
  \citet[][\S 3]{wang2017falsification} show that these restrictions may be tested by applying a modified Gail-Simon test for interaction after re-coding the data. Applying this test to the data considered by \cite{abadie2003semiparametric} confirms that they are compatible with A1 and A2 at $\alpha-$level 0.05. 
  The Gail-Simon test was performed conditional on the discrete covariates: {\sc family income}\; \$20,000, \$20,000-30,000, \$30,000-40,000, \$40,000-50,000, \$50,000-75,000 and above \$75,000; {\sc age}\; 29 years old or younger, 30-35 years old, 36-44 years old, 45-54 years old, and 55 years old or older; a {\sc marriage} indicator.  
  
  In the following, we use \cite{abadie2003semiparametric}'s instrumental variable model to estimate  the multiplicative local average treatment effect of  401(k) participation on the probability of holding an Individual Retirement Accounts. Figure S2 in the Supplementary Material gives a graphical representation of the instrumental variable model assumed in our analysis.  Throughout  we make the following assumption on the local average treatment effect:
  \begin{equation}
  \label{eqn:main_mlate}
  	\MLATE(X) =  \exp(\alpha^\top X),
  \end{equation}
 where the covariates $X$ include an intercept, family income, $\text{family income}$ squared, age,  marital status and family size. Since there are no defiers or always takers, the multiplicative local average treatment effect can also be interpreted as the multiplicative treatment effect of 401(k) participation among those who actually participated in 401(k) plans.  
%Following \cite{chernozhukov2004effects} and \cite{conley2012plausibly}, we control for income using  categorical variables: $<$\$10K, \$10-20K, \$20-30K, \$30-40K, \$40-50K, \$50-75K, and \$75K+, and control for age using categorical variables: less than 30 years old, 30-35
We apply the following estimation methods previously evaluated in the simulations: {\tt mle}, {\tt drw},  {\tt drw.ogburn}, {\tt dru.ogburn}, {\tt dru.simple}, {\tt ls.abadie} and {\tt mle.crude}. 
Since only eligible individuals may participate in 401(k) plans, one can show that  $E(H\mid X) = E(Y\mid Z=0,X)$  and $\phi_2(X)=0$, where $H=H(Y,D,X)$. Similar to the proof of Theorem \ref{thm:para}, one can show that in this situation, the models of $P(D=d,Y=y\mid Z=z,X)$, $d,y,z\in\{0,1\}$ can be  determined by the models of $\MLATE(X), \phi_1(X), \phi_3(X)$ and $OP^{CO}(X)$. 
  We provide   details of model specifications in  Section 5.5 of the Supplementary Material.
 The confidence intervals are obtained based on 500 bootstrap samples.

\begin{figure}
	\centering
	\includegraphics[width = \textwidth]{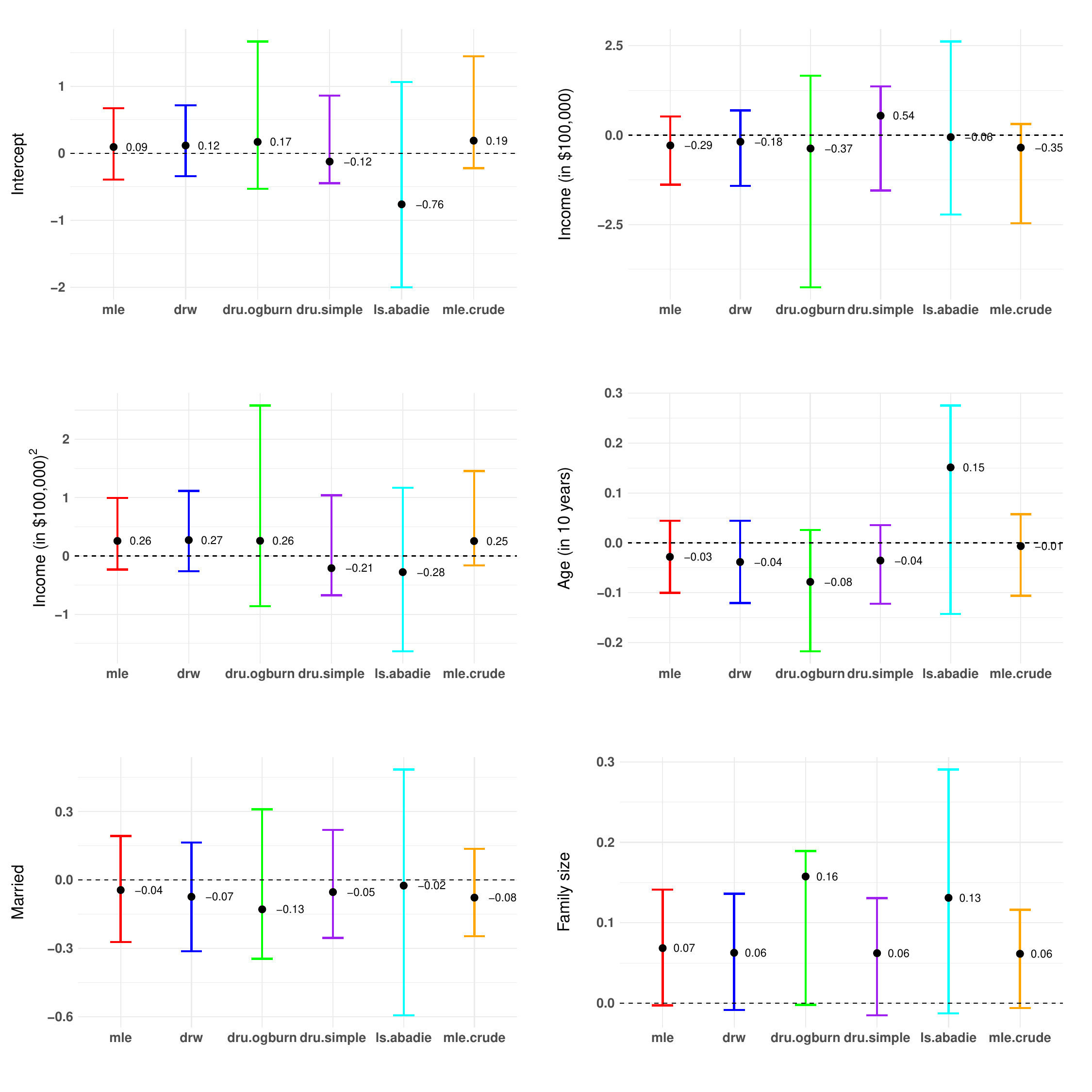}
	\caption{\hlc{Comparison of estimates of coefficients in the multplicative local average treatment effect model \eqref{eqn:main_mlate} obtained using different methods. The black dots correspond to the point estimates, and the line segments correspond to the associated 95\% confidence intervals.}
}
	\label{fig:coef}
\end{figure}

Figure \ref{fig:coef} compares coefficient estimates  for model \eqref{eqn:main_mlate}.  For example, results by {\tt mle} suggests that with each additional family member, the multiplicative effect of 401(k) participation on holding an IRA account increases by ${\rm exp}(0.068)-1 = 7.0\%$ (95\% CI = [-0.3\%, 15.1\%]). 
Results for {\tt drw.ogburn} are not plotted as its variance is huge compared to the other estimators.  This suggests that the model for $E(DY\mid Z=1,X)$ and/or the model for $E[\{H-E(H\mid X)\}^2/f^2(Z\mid X)\mid X]$ are probably mis-specified. 
% From {\tt drw}, {\tt dru.ogburn}, {\tt dru.simple} and {\tt ls.abadie}, we can get similar conclusions. All of them suggest that no effect of covariate is significant at the significance level of $\alpha=0.05$. Thus, participation in 401(k) seems to have a small or null effect on the probability of holding an IRA account. These are consistent with the observations by \cite{abadie2003semiparametric}. 
For the rest, the 95\%  confidence intervals obtained using {\tt drw} are narrower than those obtained using {\tt dru.ogburn} and {\tt dru.simple}.  This suggests that adopting the optimal weighting function is useful for reducing the variability of effect estimates. None of the covariates considered here is a significant modifier for the crowding out effect of the 401(k) plan at $\alpha-$level 0.05. 
% Although the 95\%  confidence intervals obtained using {\tt mle} are narrower than those obtained using other methods,  {\tt mle} tends to suggest that family size is a potential modifier for the crowding out effect of 401(k) plans, which is different from that of other methods,
% The results of {\tt mle.crude} is obviously different from that of other methods, it implies that there may exist unmeasured confounding. 
% The confidence intervals obtained using  {\tt dru.ogburn} and {\tt ls.abadie} are wider than the others. 
% Besides, different from other methods, the effects of all the covariates based on {\tt ls.abadie} are not significant, which means that the working models assumed for {\tt ls.abadie} are likely to be mis-specified. 

\begin{figure}
	\centering
	\includegraphics[width = \textwidth]{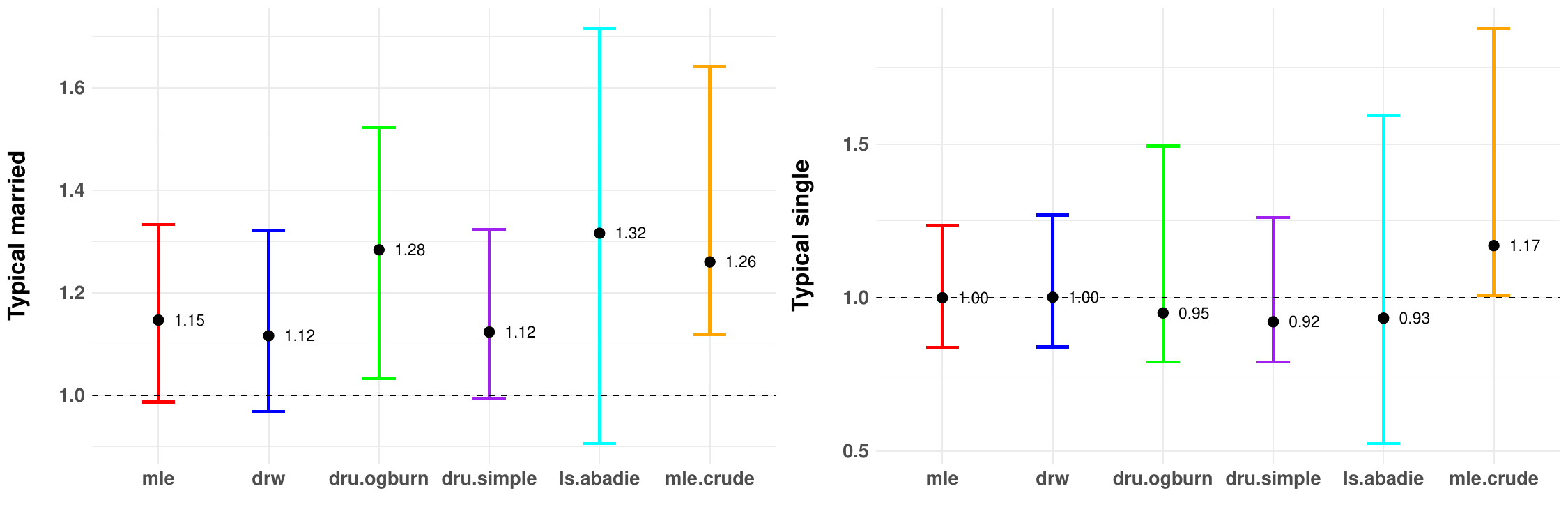}
	\caption{Comparison of estimated multiplicative local average treatment effects within two representative subgroups. The black dots correspond to the point estimates of the local average treatment effect for a typical married subject (the left panel) or a typical unmarried subject (the right panel). The line segments correspond to the associated 95\% confidence intervals.
	}
	\label{fig:subgroup}
\end{figure}

We also examine representative subgroups for married and unmarried individuals and present the results in Figure \ref{fig:subgroup}. The typical married subjects in this data set, defined by the median of individual covariates, were 40 years old, had an annual income of \$40530 and a family of size of 4. Correspondingly, the typical unmarried subjects in this data set were 39 years old, had an annual income of \$23718 and no other family members. 
Analysis results from {\tt mle} suggest that for a typical married subject, participation in the 401(k) program increases the likelihood of holding an IRA account by 14.7\% (95\% CI = [-1.2\%, 33.3\%]). In comparison, participation in the 401(k) program has virtually no effect on holding an IRA account for a typical unmarried subject. 
In either case, there is no evidence for the  crowding out effect  of 401(k) participation.
Comparing the results of {\tt mle.crude} with {\tt mle} and {\tt drw}, Figure \ref{fig:subgroup} shows that the instrumental variable methods attenuate the estimated effect of 401(k) participation on the probability of holding an Individual Retirement Account.  These findings  are consistent with the observations by \cite{abadie2003semiparametric}.

\section*{Acknowledgments}

The authors thank Elizabeth Ogburn for helpful conversations, and the referees and the associate editor for their insightful comments.  This research was supported by grants from Natural Sciences and Engineering Research Council of Canada,  U.S. National Institutes of Health and Office of Naval Research.

	\thispagestyle{empty}
	\bibliographystyle{apalike}
	\bibliography{Causal}

\clearpage

 \centerline{\large\bf ``Supplementary Material for ``Estimation of local treatment}
\vspace{2pt}
 \centerline{\large\bf under the binary instrumental variable model''}
\vspace{.25cm}
\vspace{.4cm} \centerline{Linbo Wang, Yuexia Zhang, Thomas S. Richardson, James M. Robins} \vspace{.4cm}  \vspace{.55cm} 
% \fontsize{9}{11.5pt plus.8pt minus
% .6pt}\selectfont
\par

\setcounter{equation}{0}
\setcounter{figure}{0}
\setcounter{table}{0}
\setcounter{section}{0}
\renewcommand{\theequation}{S\arabic{equation}}
\renewcommand{\thefigure}{S\arabic{figure}}
\renewcommand{\thetable}{S\arabic{table}}
\renewcommand{\theassumption}{S\arabic{assumption}}
\def\thesection{S\arabic{section}}

	\begin{abstract}
In the Supplementary Material we provide proofs of theorems and claims in the main paper. We also provide additional details for the simulations and data application.
	\end{abstract}

% 	\appendix
% 	\section*{Appendix}
% 	\renewcommand{\thesection}{\Alph{section}}

	\section{Proof of Theorem 1}
	\label{subsec:proof1}
	\begin{proof}
Under the principal stratum framework, the population can also be divided into four strata based on values of $(Y(1), Y(0))$ as in Table  \ref{table:defY}. For simplicity, we denote the principal stratum based on values of $(Y(1), Y(0))$ as $t_Y$.		

\begin{table}[!ht]
	\centering
	\caption{Principal stratum $t_Y$ based on $(Y(1),Y(0))$  }\label{table:defY}
	\begin{tabular}{cc|cc}
		\toprule
		$Y(1)$ & $Y(0)$ & Principal stratum & Abbreviation \\
		\midrule
		1 & 1 &  Always recovered & AR \\
		1 & 0 &  Helped  & HE \\
		0 & 1 &  Hurt & HU \\
		0 & 0 &  Never recovered & NR \\
		\bottomrule
	\end{tabular}
\end{table}

		We first show that (5) is a well-defined map. This follows since $\theta(X)$ is identifiable  following (1), (2) and
		\begin{flalign*}
			\phi_1(X) &= 1 - p_X(0,0\mid 1) - p_X(0,1\mid 1) - p_X(1,0\mid 0) - p_X(1,1\mid 0); \\
				\phi_2(X) &= \{p_X(1,0\mid 0)+ p_X(1,1\mid 0)\} / \{  p_X(0,0\mid 1)+ p_X(0,1\mid 1) +    p_X(1,0\mid 0)+ p_X(1,1\mid 0)\}; \\
			\phi_3(X) &= p_X(0,1\mid 1) / \{ p_X(0,0\mid 1)+ p_X(0,1\mid 1)  \}; \\
				\phi_4(X) &= p_X(1,1\mid 0) / \{ p_X(1,0\mid 0)+ p_X(1,1\mid 0)  \}; \\[5pt]
			OP^{CO}(X) &= \dfrac{E\{Y(1)I(t_D = CO)\mid X\}  E\{Y(0)I(t_D = CO)\mid X\} }{[p(CO;X)-E\{Y(1)I(t_D = CO)\mid X\} ] [p(CO;X)-E\{Y(0)I(t_D = CO)\mid X\}] } \\
			&= \dfrac{\{p(CO,HE;X) + p(CO,AR;X)\}   \{p(CO,HU;X) + p(CO,AR;X)\}  } 
			{    {\{p(CO,NR;X) + p(CO,HU;X)\}   \{p(CO,NR;X) + p(CO,HE;X)\}  }   } \\
			&= \dfrac{ \{p_X(1,1\mid 1)-p_X(1,1\mid 0)\}   \{p_X(0,1\mid 0)-p_X(0,1\mid 1)\}   } 
			{   \{p_X(1,0\mid 1)-p_X(1,0\mid 0)\}   \{p_X(0,0\mid 0)-p_X(0,0\mid 1)\}    }
		\end{flalign*}
		are identifiable from $\bm p = (p_x(d,y\mid z); d,y,z=0,1).$ 
		
		 To show that (5) is a bijection, for each realization of $X$, let $\bm c = (c_0,\ldots,c_5)$ be a vector in $\mathcal{D} \times [0,1]^4 \times [0,+\infty).$  We need to show there is one and only one $\bm p \in \Delta$ such that 
		 \begin{equation}
		 \label{eqn:2-5}
		 	(\phi_1(X), \ldots, \phi_4(X)) = (c_1,\ldots, c_4)
		 \end{equation}
		 and
		 \begin{equation}
		 \label{eqn:16}
		 	\left(\theta(X), OP^{CO}(X)\right) = (c_0,c_5).
		 \end{equation}
		 For  simplicity of notation, we suppress the dependence on $X$ in the remainder of the proof.
	First note that \eqref{eqn:2-5} implies that
	\begin{flalign*}
		p(0,1\mid 1) = (1-c_1)(1-c_2)c_3; &\quad p(0,0\mid 1) =  (1-c_1)(1-c_2)(1-c_3); \\
		p(1,1\mid 0) = (1-c_1)c_2c_4; &\quad p(1,0\mid 0) =  (1-c_1)c_2(1-c_4). \numberthis \label{eqn:7}    
	\end{flalign*}	 
	According to \cite{richardson2017modeling}, \eqref{eqn:16} implies that
	\begin{flalign*}
		E\{Y(1)\mid  t_D = CO\} &= \dfrac{p(1,1\mid 1) - p(1,1\mid 0)}{p(CO)}  = f_1(c_0,c_5); \\
			E\{Y(0)\mid  t_D = CO \} &= \dfrac{p(0,1\mid 0) - p(0,1\mid 1)}{p(CO)}  = f_0(c_0,c_5),  \numberthis \label{eqn:8}  
	\end{flalign*}
	where $p(CO) = 1-p(0,0\mid 1) - p(0,1\mid 1) - p(1,0\mid 0) - p(1,1\mid 0)$ and $f_d(c_0,c_5), d=0,1$ are known smooth functions of $c_0,c_5$ that take values between 0 and 1. The functional form of $f_d(c_0, c_5)$ follows from equations (2.4) and (2.5) in \cite{richardson2017modeling}. Specifically, 
	\[
	f_0(c_0,c_5)=
	\begin{cases}
	\dfrac{1}{2(c_5-1)} \left\{ 
c_5(2-c_0)+c_0-\sqrt{c_0^2(c_5-1)^2+4c_5}\right\} &\theta = \LATE;\\
	\dfrac{1}{2c_0(1-c_5)} \left\{ 
	-(c_0+1)c_5+\sqrt{c_5^2(c_0-1)^2 +4c_0c_5}\right\} & \theta = \MLATE;
	\end{cases}
	\]
	and
		\[
	f_1(c_0,c_5)=
	\begin{cases}
	f_0(c_0,c_5)+c_0 &\theta = \LATE;\\
	f_0(c_0,c_5)c_0 & \theta = \MLATE.
	\end{cases}
	\]

	Combining \eqref{eqn:7} and \eqref{eqn:8},  we have \eqref{eqn:2-5} and \eqref{eqn:16} together imply  that 
	\begin{flalign*}
		p(0,1\mid 1) = (1-c_1)(1-c_2)c_3; &\quad p(0,0\mid 1) =  (1-c_1)(1-c_2)(1-c_3); \\
	p(1,1\mid 0) = (1-c_1)c_2c_4; &\quad p(1,0\mid 0) =  (1-c_1)c_2(1-c_4); \\
		p(1,1\mid 1) &= f_1(c_0,c_5) c_1 + p(1,1\mid 0); \\
		p(0,1\mid 0) &= f_0(c_0,c_5) c_1 + p(0,1\mid 1); \\ \numberthis \label{eqn:10}    
		p(1,0\mid 1) &= 1-p(0,0\mid 1)  - p(0,1\mid 1) - p(1,1\mid 1); \\
		p(0,0\mid 0) &= 1-p(0,1\mid 0) - p(1,0\mid 0) - p(1,1\mid 0).
	\end{flalign*}

	 We now only need to show $\bm p$ defined in \eqref{eqn:10} lies in $\Delta.$ First note that 
	\begin{flalign*}
	p(1,1\mid 1) &= f_1(c_0,c_5) c_1 + p(1,1\mid 0) 
	\leq c_1 + (1-c_1)c_2c_4 \leq 1; \\
	p(0,1\mid 0) &= f_0(c_0,c_5) c_1 + p(0,1\mid 1) \leq c_1 + (1-c_1)(1-c_2)c_3 \leq 1; \\
	p(1,0\mid 1) &= 1 - p(0,0\mid 1)  - p(0,1\mid 1) - p(1,1\mid 1)\\
	 &=1- (1-c_1)(1-c_2) - f_1(c_0,c_5) c_1 - p(1,1\mid 0) \\
		&\geq  1-(1-c_1)(1-c_2) - c_1  - (1-c_1)c_2c_4 \\
		&= (1-c_1)c_2(1-c_4) \geq 0; \numberthis \label{eqn:11}     \\
	p(0,0\mid 0) &= 1-p(0,1\mid 0) - p(1,0\mid 0) - p(1,1\mid 0) \\
		 &= 1-f_0(c_0,c_5)c_1 - p(0,1\mid 1)  - (1-c_1)c_2 \\
		&\geq  1- c_1  -  (1-c_1)(1-c_2)c_3 - (1-c_1)c_2   \\
		&= (1-c_1)(1-c_2)(1-c_3) \geq 0. \numberthis \label{eqn:12}    
	\end{flalign*}
	Furthermore,
	\begin{flalign*}
		p(1,1\mid 1) &= f_1(c_0,c_5) c_1 + p(1,1\mid 0) \geq p(1,1,0); \\
		p(1,0\mid 1) &\geq (1-c_1)c_2(1-c_4) = p(1,0\mid 0); \\
		p(0,1\mid 0) &= f_0(c_0,c_5) c_1 + p(0,1\mid 1) \geq p(0,1\mid 1); \\
		p(0,0\mid 0) &\geq (1-c_1)(1-c_2)(1-c_3) = p(0,0\mid 1),
 	\end{flalign*}
 	where the second and last inequality were shown in \eqref{eqn:11} and \eqref{eqn:12}.
	
%	Finally, 	\begin{flalign*}
%			p(1,0\mid 1) + p(1,1\mid 0) &\leq 1-(1-c_1)(1-c_2)-c_1= (1-c_1)c_2 \leq 1; \\
%			p(1,1\mid 1) + p(1,0\mid 0) &\leq c_1 + (1-c_1)c_2c_4  + (1-c_1)c_2(1-c_4) = c_1 + (1-c_1)c_2 \leq 1; \\
%			p(0,1\mid 1) + p(0,0\mid 0) &\leq 1-c_1 - (1-c_1)c_2 = (1-c_1)(1-c_2) \leq 1;\\
%			p(0,0\mid 1) + p(0,1\mid 0) &\leq (1-c_1) (1-c_2)(1-c_3) + c_1 + (1-c_1) (1-c_2)c_3 \\ &= c_1 + (1-c_1) (1-c_2) \leq 1,
%	\end{flalign*}
%		where the first and third inequality follow from \eqref{eqn:11} and \eqref{eqn:12}.
%	
	We have hence finished the proof.
	\end{proof}

\section{Proof of Theorem 2}
	\begin{proof}
	
	If $\theta(X) = \LATE(X)$, then $H(Y,D,X)=Y-D\theta(X)$.	Since
	\[
	\delta^L(X) = \dfrac{E(Y\mid Z=1,X) - E(Y\mid Z=0,X)}{E(D\mid Z=1,X) - E(D\mid Z=0,X)},
	\]
	then 
	
	\begin{align*}
	\MoveEqLeft{E\{H(Y,D,X)\mid Z=1,X\}-E\{H(Y,D,X)\mid Z=0,X\}}\\
&= 	E(Y\mid Z=1,X)-E(D\mid Z=1,X)\theta(X)-E(Y\mid Z=0,X)+E(D \mid Z=0,X)\theta(X)\\[2pt]
&=E(Y\mid Z=1,X)-E(Y\mid Z=0,X)\\[-4pt]
&\quad -\big\{E(D\mid Z=1,X)
-E(D\mid Z=0,X)\big\}\dfrac{E(Y\mid Z=1,X) - E(Y\mid Z=0,X)}{E(D\mid Z=1,X) - E(D\mid Z=0,X)}=0.
	\end{align*}
	
	Thus, $E\{H(Y,D,X)\mid X\}=E\{H(Y,D,X)\mid Z=0,X\}$.
Based on the results in Section \ref{subsec:proof1}, we have 

\begin{align*}
\MoveEqLeft{E\{H(Y,D,X)\mid X\}=E\{H(Y,D,X)\mid Z=0,X\}}\\
&=E(Y\mid Z=0,X)-E(D\mid Z=0,X)\theta(X)\\
&=P(Y=1\mid Z=0,X)-P(D=1\mid Z=0,X)\theta(X)\\
&=P(D=1,Y=1\mid Z=0,X)+P(D=0,Y=1\mid Z=0,X)-P(D=1\mid Z=0,X)\theta(X)\\
&=p_X(1,1\mid 0)+p_X(0,1\mid 0)-\theta(X)\{1-\phi_1(X)\}\phi_2(X)\\
&=\{1-\phi_1(X)\}\phi_2(X)\phi_4(X)+f_0\{\theta(X),OP^{CO}(X)\}\phi_1(X)\\
&\quad +\{1-\phi_1(X)\}\{1-\phi_2(X)\}\phi_3(X)-\theta(X)\{1-\phi_1(X)\}\phi_2(X).
\end{align*}

If $\theta(X) = \MLATE(X)$, then $H(Y,D,X)=Y\theta(X)^{-D}$. Since
$$
\delta^M(X) = -\dfrac{E(YD\mid Z=1,X) - E(YD\mid Z=0,X)}{E\{Y(1-D)\mid Z=1,X\} - E\{Y(1-D)\mid Z=0,X\}},
$$
then 
	\begin{align*}
\MoveEqLeft{E\{H(Y,D,X)\mid Z=1,X\}-E\{H(Y,D,X)\mid Z=0,X\}}\\
&= E\{Y\theta(X)^{-D}\mid Z=1,X\}-E\{Y\theta(X)^{-D}\mid Z=0,X\}\\
&=\theta(X)^{-1}P(Y=1,D=1\mid Z=1,X)+P(Y=1,D=0\mid Z=1,X)\\
&\quad -\theta(X)^{-1}P(Y=1,D=1\mid Z=0,X)-P(Y=1,D=0\mid Z=0,X)\\
&=\theta(X)^{-1}\big\{P(Y=1,D=1\mid Z=1,X)-P(Y=1,D=1\mid Z=0,X)\big\}\\
&\quad +P(Y=1,D=0\mid Z=1,X)-P(Y=1,D=0\mid Z=0,X)\\
&=-\dfrac{P(Y=1,D=0\mid Z=1,X)-P(Y=1,D=0\mid Z=0,X)}{P(Y=1,D=1\mid Z=1,X)-P(Y=1,D=1\mid Z=0,X)}\\
&\quad\quad \times \big\{P(Y=1,D=1\mid Z=1,X)-P(Y=1,D=1\mid Z=0,X)\big\}\\
&\quad +P(Y=1,D=0\mid Z=1,X)-P(Y=1,D=0\mid Z=0,X)\\
&=0.
	\end{align*}
	Thus, $E\{H(Y,D,X)\mid X\}=E\{H(Y,D,X)\mid Z=0,X\}$. Based on the results in Section \ref{subsec:proof1}, we have 
	\begin{align*}
\MoveEqLeft{E\{H(Y,D,X)\mid X\}=E\{H(Y,D,X)\mid Z=0,X\}}\\
	&=E\{Y\theta(X)^{-D}\mid Z=0,X\}\\
	&=\theta(X)^{-1}P(Y=1,D=1\mid Z=0,X)+P(Y=1,D=0\mid Z=0,X)\\
	&=\{1-\phi_1(X)\}\phi_2(X)\phi_4(X)\theta(X)^{-1}\\
	&\quad +f_0\{\theta(X),OP^{CO}(X)\}\phi_1(X)+\{1- \phi_1(X)\}\{1- \phi_2(X)\}\phi_3(X).
	\end{align*}
	
Therefore, $\hat{E}\{H(Y,D,X;\alpha)\mid X\}$ has the form as shown in Theorem 2.

If the model for $E\{H(Y,D,X;\alpha)\mid X\}$ is correctly specified, but the model for $f(Z\mid X;\gamma)$ may be mis-specified, then
 $\hat{\gamma}\stackrel{p}\rightarrow \gamma^\ast$ under some regularity conditions \citep{white1982maximum}, where $\gamma^\ast$ is not necessarily equal to  $\gamma$. Furthermore, 
\[
\begin{aligned}
&E\left[\mathbb{P}_n\omega(X) \dfrac{2Z-1}{f(Z\mid X;\gamma^\ast)} \left[H(Y,D,X;\alpha) - E\left\{ H(Y,D,X;\alpha) \mid Z=0,X\right\} \right]\right]\\
=&E\left[E\left\{\mathbb{P}_n\omega(X) \dfrac{2Z-1}{f(Z\mid X;\gamma^\ast)} \left[H(Y,D,X;\alpha) - E\left\{ H(Y,D,X;\alpha) \mid Z=0,X\right\} \right]\Big| Z=0,X\right\}\right]\\
=&E\left[ \mathbb{P}_n\omega(X) \dfrac{-1}{f(0\mid X;\gamma^\ast)} \left[E\left\{ H(Y,D,X;\alpha) \mid Z=0,X\right\} - E\left\{ H(Y,D,X;\alpha) \mid Z=0,X\right\} \right]\right]\\
=&0.
\end{aligned}
\]

If the model for $f(Z\mid X;\gamma)$ is correctly specified, but the model for $E\{H(Y,D,X;\alpha)\mid X\}$ may be mis-specified, then   $\hat{E}\{H(Y,D,X;\alpha)\mid X\}-E\{H(Y,D,X;\alpha)\mid X\}$  does not necessarily converge to zero in probability. Assume there exists $E^{*}\{H(Y,D,X;\alpha)\mid X\}$ such that $\hat{E}\{H(Y,D,X;\alpha)\mid X\}-E^{*}\{H(Y,D,X;\alpha)\mid X\}\stackrel{p}\rightarrow 0$ under some regularity conditions.  Furthermore, 

\begin{align*}
\MoveEqLeft{E\left[\mathbb{P}_n \omega(X) \dfrac{2Z-1}{f(Z\mid X;\gamma)} \left[H(Y,D,X;\alpha) - E^{*}\left\{ H(Y,D,X;\alpha) \mid X\right\} \right]\right]}\\
&=E\left[\mathbb{P}_n \omega(X) \dfrac{2Z-1}{f(Z\mid X;\gamma)}H(Y,D,X;\alpha)\right]-E\left[\mathbb{P}_n \omega(X) \dfrac{2Z-1}{f(Z\mid X;\gamma)}E^{*}\left\{ H(Y,D,X;\alpha) \mid X\right\}\right]\\
&=E\left[E\left\{\mathbb{P}_n \omega(X) \dfrac{2Z-1}{f(Z\mid X;\gamma)}H(Y,D,X;\alpha)\Big| X\right\}\right]\\
&\quad-E\left[E\left\{\mathbb{P}_n \omega(X) \dfrac{2Z-1}{f(Z\mid X;\gamma)}E^{*}\left\{ H(Y,D,X;\alpha) \mid  X\right\}\Big| X\right\}\right]\\
&=E\left[E\left\{\mathbb{P}_n \omega(X) \dfrac{2Z-1}{\{P(Z=1\mid X;\gamma)\}^Z\{1-P(Z=1\mid X;\gamma)\}^{1-Z}}H(Y,D,X;\alpha)\Big| X\right\}\right]\\
&\quad-E\left[\mathbb{P}_n \omega(X)E^{*}\left\{ H(Y,D,X;\alpha) \mid X\right\}E\left\{ \dfrac{2Z-1}{\{P(Z=1\mid X;\gamma)\}^Z\{1-P(Z=1\mid X;\gamma)\}^{1-Z}}\Big| X\right\}\right]\\
&=E\Big[\mathbb{P}_n \omega(X)\big[E\{H(Y,D,X;\alpha)\mid Z=1,X\}-E\{H(Y,D,X;\alpha)\mid Z=0,X\}\big]\Big]-0\\
&=0.
\end{align*}

The rest of proof follows from standard M-estimation theory.
	
	\end{proof}
	
	\section{Optimal weighting function}
	\label{appendix:optimal}
	
	If $\theta(X) = \LATE(X)$, \cite{ogburn2015doubly} show that the optimal choice of $\omega(X)$ is given by
	\begin{flalign*}
		\omega_{\rm opt}(X) &= -\dfrac{\partial \theta(X)}{\partial \alpha} 
		E\left\{ \left. \dfrac{2Z-1}{f(Z\mid X)} D\right| X \right\} 
		E^{-1} \left[\left. \dfrac{\left\{H - E(H\mid Z,X)\right\}^2}{f^2(Z\mid X)} \right| X \right] \\
		&=  -\dfrac{\partial \theta(X)}{\partial \alpha} \times \phi_1(X) \times \left[   E_{Z\mid X} \dfrac{1}{f^2(Z\mid X)} \left \{  E\left(H^2\mid Z,X\right)  -  E^2\left(H\mid Z,X\right) \right\}        \right]^{-1},
	\end{flalign*}
with
\[
	H=H(Y,D,X)=Y-D\theta(X);	
\]
		\begin{align*}
\MoveEqLeft{E\left(H^2\mid Z=z,X\right) = E\left[\{Y-D \theta(X)\}^2\mid Z=z,X  \right]}\\
%	&= E\left(Y^2 - 2DY\theta + D^2 \theta^2 \mid Z=1,X \right) \\
	&= P(Y=1\mid Z=z,X) + \theta(X)^2 P(D=1\mid Z=z,X)  - 2 \theta(X)  P(DY=1\mid Z=z,X);\\
%	&= p(0,1\mid z) + p(1,1\mid z) + \theta^2 \left\{ p(1,1\mid z) + p(1,0\mid z) \right\} -   2\theta p(1,1\mid z)\\
%		E\left(H^2\mid Z=1,X\right) &=(1-\phi_1)(1-\phi_2)\phi_3 + f_1 \phi_1+ (1-\phi_1)\phi_2\phi_4 +  \theta^2 \{ 1-(1-\phi_1)(1-\phi_2) \} - \\
%		& 2\theta \{f_1c_1 +(1-\phi_1) \phi_2\phi_4 \}  \\
%	E\left(H^2\mid Z=0,X\right) &= f_0 \phi_1 + (1-\phi_1)(1-\phi_2)\phi_3 + (1-\phi_1)\phi_2\phi_4 +  \theta^2 (1-\phi_1) \phi_2 -  2\theta (1-\phi_1) \phi_2\phi_4 \\
\end{align*}
\begin{align*}
\MoveEqLeft{E(H\mid Z,X) = E\{Y-D\theta(X) \mid Z=0,X\}} \\
	&= P(Y=1\mid Z=0,X) - \theta(X) P(D=1\mid Z=0,X),
%	&= f_0 \phi_1 + (1-\phi_1)(1-\phi_2)\phi_3 + (1-\phi_1) \phi_2 \phi_4 - \theta (1-\phi_1)\phi_2 \\
	\end{align*}
where as shown in Section \ref{subsec:proof1}, $P(Y=1\mid Z=z,X), P(D=1\mid Z=z,X), P(DY=1\mid Z=z, X)$ are functions of $\{\theta(X), \phi_1(X),\ldots,\phi_4(X), OP^{CO}(X)\}.$

If $\theta(X) = \MLATE(X)$, \cite{ogburn2015doubly} show that the optimal choice of $\omega(X)$ is given by
\begin{flalign*}
\omega_{\rm opt}(X) &= -\dfrac{\partial \theta(X)}{\partial \alpha}  \theta(X)^{-2} 
E\left\{ \left. \dfrac{2Z-1}{f(Z\mid X)} D Y \right| X \right\} 
E^{-1} \left[\left. \dfrac{\left\{H - E(H\mid Z,X)\right\}^2}{f^2(Z\mid X)} \right| X \right] \\
&=  -\dfrac{\partial \theta(X)}{\partial \alpha} \times \theta(X)^{-2} \times f_1\phi_1(X) \times \left[   E_{Z\mid X} \dfrac{1}{f^2(Z\mid X)} \left \{  E\left(H^2\mid Z,X\right)  -  E^2\left(H\mid Z,X\right) \right\}        \right]^{-1},
\end{flalign*}
with
\begin{flalign*}
H&=H(Y,D,X)=Y\theta(X)^{-D}; \quad f_1=f_1\{\theta(X),OP^{C0}(X)\};\\
E\left(H^2\mid Z=z,X\right) &= E\left\{Y^2 \theta(X)^{-2D}\mid Z=z,X  \right\} \\
%	&= E\left(Y^2 - 2DY\theta + D^2 \theta^2 \mid Z=1,X \right) \\
&= P(Y=1,D=1\mid Z=z,X) \theta(X)^{-2} + P(Y=1,D=0\mid Z=z,X);\\
%	&= p(0,1\mid z) + p(1,1\mid z) + \theta^2 \left\{ p(1,1\mid z) + p(1,0\mid z) \right\} -   2\theta p(1,1\mid z)\\
%		E\left(H^2\mid Z=1,X\right) &=(1-\phi_1)(1-\phi_2)\phi_3 + f_1 \phi_1+ (1-\phi_1)\phi_2\phi_4 +  \theta^2 \{ 1-(1-\phi_1)(1-\phi_2) \} - \\
%		& 2\theta \{f_1c_1 +(1-\phi_1) \phi_2\phi_4 \}  \\
%	E\left(H^2\mid Z=0,X\right) &= f_0 \phi_1 + (1-\phi_1)(1-\phi_2)\phi_3 + (1-\phi_1)\phi_2\phi_4 +  \theta^2 (1-\phi_1) \phi_2 -  2\theta (1-\phi_1) \phi_2\phi_4 \\
E(H\mid Z,X) &= E\{Y\theta(X)^{-D} \mid Z=0,X\} \\
&= P(Y=1,D=1\mid Z=0,X)\theta(X)^{-1} + P(Y=1,D=0\mid Z=0,X).
%	&= f_0 \phi_1 + (1-\phi_1)(1-\phi_2)\phi_3 + (1-\phi_1) \phi_2 \phi_4 - \theta (1-\phi_1)\phi_2 \\
\end{flalign*}

\section{Proof of the variation independence claims in Remark 2}

We first show the variation independence between $E(Y\mid X), E(D\mid X), f(Z\mid X)$ and $\delta^L(X).$ For a particular realization of $X$, and any  $(a,b,\pi,c) \in (0,1)^3 \times (-1,1),$ we need to show  it is possible that 
\begin{flalign}
    E(Y\mid X) &= a;\label{eqn:y} \\
    E(D\mid X) &= b;\label{eqn:d} \\
    P(Z=1\mid X) &= \pi; \label{eqn:pi}\\
    \delta^L(X) &= c. \nonumber
\end{flalign}
It is clear that \eqref{eqn:y} -- \eqref{eqn:pi} may hold simultaneously since $E(Y\mid X), E(D\mid X)$ and $f(Z\mid X)$ are clearly variation independent. Equations \eqref{eqn:y} -- \eqref{eqn:pi} place the following constraints on the range of $a_1 \equiv E(Y\mid Z=1,X), a_0 \equiv E(Y\mid Z=0,X), b_1\equiv E(D\mid Z=1,X), b_0 \equiv E(D\mid Z=0,X)$:
\begin{flalign}
    \pi a_1 + (1-\pi) a_0 &= a;\label{eqn:lp1} \\
    \pi b_1 + (1-\pi) b_0 &= b;\label{eqn:lp2} \\
    0\leq a_1, a_0, b_1, b_0 &\leq 1. \label{eqn:lp3}
\end{flalign}
We use a  linear programming algorithm to obtain the range of  $\delta^Y = a_1 - a_0$ and $\delta^D = b_1-b_0$ subject to the constraints in  \eqref{eqn:lp1}--\eqref{eqn:lp3}:
\begin{flalign*}
    {\rm max}\left(-\dfrac{a}{1-\pi}, -\dfrac{1-a}{\pi}\right) &\leq \delta^Y \leq {\rm min}\left( \dfrac{1-a}{1-\pi}, \dfrac{a}{\pi}  \right); \\
    {\rm max}\left(-\dfrac{b}{1-\pi}, -\dfrac{1-b}{\pi}\right) &\leq \delta^D \leq {\rm min}\left( \dfrac{1-b}{1-\pi}, \dfrac{b}{\pi}  \right).
\end{flalign*}
Note that as long as $0<a,b,\pi<1$, the feasible range of $(\delta^Y, \delta^D)$ always contains a ball around the origin. Hence the range of $c = \delta^Y / \delta^D$ is unrestricted by the constraints in  \eqref{eqn:y} -- \eqref{eqn:pi}.

The proof of the variation independence among $E(Y\mid X), E(Y\mid D, X), f(Z\mid X)$ and $\delta^M(X)$ follows the same logic and is hence omitted.

\section{Additional details for simulation studies and data analysis}

\subsection{Visualization of the degree of model mis-specification}

\begin{figure}[!htbp]
	\centering
	\includegraphics[width=32em,angle=0]{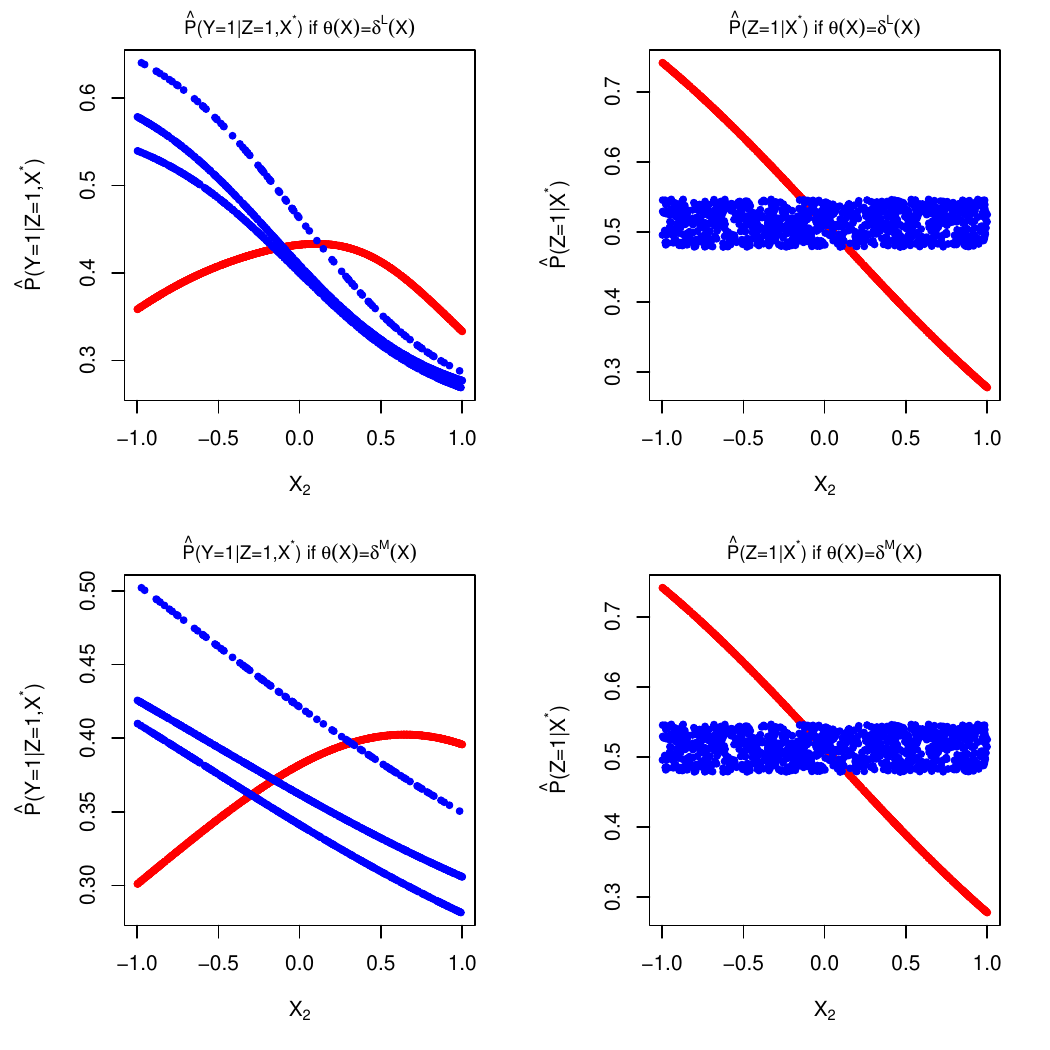}
	\caption{Visualizing nuisance model mis-specification with
	correct specification for $\theta(X)$: top row: $\theta(X) =  \LATE(X)$; bottom row: $\theta(X) = \MLATE(X)$; red dots: correct nuisance model specification; blue dots: incorrect nuisance model specification
}
	\label{fig:misspecify}
\end{figure}

In Figure \ref{fig:misspecify}, we provide visualization of the degree of model mis-specification using data points generated from one randomly selected Monte Carlo run. We denote the first element of covariates $X$ as $X_1$, namely the intercept. We denote the second element of $X$ as $X_2$, which was generated from ${\rm Unif}(-1,1).$ Figure \ref{fig:misspecify}  shows fitted probabilities using the correct/incorrect models as functions of $X_2$. To unify notation, we use $X^\ast$ to denote the covariate used in fitting the models, which may correspond to $X, X^\dagger$ or $X^\prime$ depending on the specific scenario. If the model for the instrumental density is correctly specified, then $\hat{P}(Z=1\mid X^\ast) = \hat{P}(Z=1\mid X)$; otherwise $\hat{P}(Z=1\mid X^\ast) = \hat{P}(Z=1\mid X^\dagger)$. If the other nuisance models are correctly specified, then the fitted probabilities $\hat{P}(Y=1\mid Z=1, X^\ast)$ are derived from the fitted values of $\theta(X), \phi_i(X), i=1,\ldots, 4$ and $OP^{CO}(X)$; otherwise, $\hat{P}(Y=1\mid Z=1, X^\ast)$ are derived from the fitted values of $\theta(X), \phi_i(X^\prime), i=1,\ldots, 4$ and $OP^{CO}(X^\prime)$.

From the left panels of Figure \ref{fig:misspecify}, one can see that under correct model specification, $\hat{P}(Y=1\mid Z=1, X)$ is a {non-monotone function} of $X_2$. Under mis-specifications of nuisance models, $\hat{P}(Y=1\mid Z=1, X^\ast)$ has three clusters, as for each value of $X_2$, it is possible that $X^\prime = (1,0), (1,1)$ or $(0,1)$. The fitted values $\hat{P}(Y=1\mid Z=1, X^\ast)$ still depends on $X_2$ through the fitted values of $\theta(X)$. From the right panels of Figure \ref{fig:misspecify}, one can see that under correct model specification, $\hat{P}(Z=1\mid X)$ is an expit function of $X_2$. Under mis-specification of the  instrumental density model, the fitted values $\hat{P}(Z=1\mid X^\dagger)$ do not depend on $X_2$ as $X^\dagger$ is independent of $X_2$. 

\subsection{Implementation details in the simulation studies}
\label{ss:imsimu}
We now describe the implementation details of the various estimators considered in the simulation studies:
\begin{itemize}
\item[\tt reg.ogburn]
\item[\tt dru.ogburn]
\item[\tt drw.ogburn]
If $\theta(X)=\LATE(X)$, then we assume  
\begin{flalign*}
P(Z=1\mid X) &= {\rm expit}(\gamma^\top X);\\
E(H\mid X)&=\xi_1+\xi_2 X_2+\xi_3 X_2^2 +\xi_4 X_2^3;\\
\delta^D(X)&=\tanh(\psi_1^\top X);\\
	E \left[\left. \dfrac{\left\{H - E(H\mid X)\right\}^2}{f^2(Z\mid X)} \right| X \right]  &= \zeta_1 + \zeta_2 X_2 + \zeta_3 X_2^2.
\end{flalign*}
  If $\theta(X)=\MLATE(X)$, then  we assume  
\begin{flalign*}
P(Z=1\mid X) &= {\rm expit}(\gamma^\top X);\\
E(H\mid X)&=\exp(\xi_5+\xi_6X_2+\xi_7X_2^2);\\
E(DY\mid Z,X)&={\rm expit}(\psi_2 Z+\psi_3^\top X);\\
E \left[\left. \dfrac{\left\{H - E(H\mid X)\right\}^2}{f^2(Z\mid X)} \right| X \right]  &= \exp(\zeta_4 + \zeta_5 X_2 + \zeta_6 X_2^2),
\end{flalign*}
where 
\[
H=H(Y,D,X)=
\begin{cases}
Y - D \theta(X)& \theta(X) = \LATE(X);\\
Y \theta(X)^{-D}	& \theta(X) = \MLATE(X).
\end{cases}
\]
The higher order terms are added so that these models better approximate the truth. 

In {\tt reg.ogburn}, we obtain $(\hat{\xi}_{\rm reg},\hat{\alpha}_{\rm reg})$ as the solution to  the following estimating equations:
\[
\begin{cases}
  \mathbb{P}_n\begin{pmatrix}(1,X_2,X_2^2,X_2^3)^\top\\Z[1-\{\theta(X;\alpha)\}^2](1,X_2)^\top \end{pmatrix}\{H(\alpha)-E(H\mid X;\xi)\}=0 & \theta(X) = \LATE(X);\\
    \mathbb{P}_n\begin{pmatrix}E(H\mid X;\xi)(1,X_2,X_2^2)^\top\\Z\theta(X;\alpha)(1,X_2)^\top \end{pmatrix}\{H(\alpha)-E(H\mid X;\xi)\}=0 & \theta(X) = \MLATE(X).
\end{cases}
\]

In {\tt dru.ogburn}, first, we fit the model $P(Z=1\mid X;\gamma)$  using {\tt R} package {\tt glm}.
Then, we obtain $\hat{\alpha}_{\rm dru.ogburn}$ as the solution to the estimating equation 
\[
 \mathbb{P}_n\frac{2Z-1}{f(Z\mid X;\hat{\gamma})}\left\{H(\alpha)-E(H\mid X;\hat{\xi}_{\rm reg})\right\}=0.
\]

In {\tt drw.ogburn},  first, we fit the model $P(Z=1\mid X;\gamma)$  using {\tt R} package {\tt glm}. Next, we estimate the optimal weighting function $\omega_{\rm opt}(X)$. If $\theta(X)=\LATE(X)$,
\[
\hat{\omega}_{\rm opt}(X)=-X[1-\{\theta(X;\hat{\alpha}_{\rm reg})\}^2]\delta^D(X;\hat{\psi})E^{-1} \left[\left. \dfrac{\left\{H(\hat{\alpha}_{\rm reg}) - E(H\mid X;\hat{\xi}_{\rm reg})\right\}^2}{f^2(Z\mid X;\hat{\gamma})} \right| X;\hat{\zeta} \right];
\]
if $\theta(X)=\MLATE(X)$,
\[
\begin{aligned}
\hat{\omega}_{\rm opt}(X)=&-X\{\theta(X;\hat{\alpha}_{\rm reg})\}^{-1}\left\{E(DY\mid Z=1,X;\hat{\psi})-E(DY\mid Z=0,X;\hat{\psi})\right\}\\
 &\times E^{-1} \left[\left. \dfrac{\left\{H(\hat{\alpha}_{\rm reg}) - E(H\mid X;\hat{\xi}_{\rm reg})\right\}^2}{f^2(Z\mid X;\hat{\gamma})} \right| X;\hat{\zeta} \right],
 \end{aligned}
\]
where the model  $\delta^D(X;\psi)$ is fitted using the doubly robust estimator of \cite{richardson2017modeling} and obtained  using {\tt R} package {\tt brm}, 
the model  $E[\{H(\hat{\alpha}_{\rm reg})-E(H\mid X;\hat{\xi}_{\rm reg})\}^2/f^2(Z\mid X;\hat{\gamma})\mid X;\zeta]$ is fitted using the least squares method with the restriction that $E[\{H(\hat{\alpha}_{\rm reg})-E(H\mid X;\hat{\xi}_{\rm reg})\}^2/f^2(Z\mid X;\hat{\gamma})\mid X;\hat{\zeta}]>0$, and the model $E(DY\mid Z,X;\psi)$ is fitted {\tt R} package {\tt glm}. 
Then, we  obtain $\hat{\alpha}_{\rm drw.ogburn}$ as the solution to the estimating equation 
\[
 \mathbb{P}_n\hat{\omega}_{\rm opt}(X)\frac{2Z-1}{f(Z\mid X;\hat{\gamma})}\left\{H(\alpha)-E(H\mid X;\hat{\xi}_{\rm reg})\right\}=0.
\]

\item[\tt mle.wang]
\item[\tt dru.wang]
If  $\theta(X)=\LATE(X)$, then  we  assume  
\begin{flalign*}
P(Z=1\mid X) &= {\rm expit}(\gamma^\top X);\\
\delta^D(X)&=\tanh(\lambda^\top X);\\
 {OP}^{D}(X)&\equiv \frac{P(D=1\mid Z=1,X)P(D=1\mid Z=0,X)}{P(D=0\mid Z=1,X)P(D=0\mid Z=0,X)}=\exp(\tau^\top X);\\
 {OP}^{Y}(X)&\equiv \frac{P(Y=1\mid Z=1,X)P(Y=1\mid Z=0,X)}{P(Y=0\mid Z=1,X)P(Y=0\mid Z=0,X)}=\exp(\kappa^\top X).
\end{flalign*}

In {\tt mle.wang}, first, we fit the models $\delta^D(X;\lambda)$ and  ${OP}^{D}(X;\tau)$ using the maximum likelihood estimation implemented in {\tt R} package {\tt brm}. Then, we fit the models ${OP}^{Y}(X;\kappa)$ and $\theta(X;\alpha)$
using the maximum likelihood estimation method based on the likelihood function of $Y$ conditional on $Z$, $X$ and $\delta^D(X;\hat{\lambda})$. The maximum likelihood estimator of $\alpha$ is denoted as ${\hat{\alpha}_{\rm mle.wang}}$.

In {\tt dru.wang}, first, we fit the model $P(Z=1\mid X;\gamma)$  using {\tt R} package {\tt glm}. Next, we get $\hat{E}(D\mid Z=0,X)$ and $\hat{E}(Y\mid Z=0,X)$ from $\delta^D(X;\hat{\lambda})$, ${OP}^{D}(X;\hat{\tau})$, $\theta(X;\hat{\alpha}_{\rm mle.wang})$ and  ${OP}^{Y}(X;\hat{\kappa})$ based on Proposition 2 of \cite{wang2018bounded}. Then,
we obtain $\hat{\alpha}_{\rm dru.wang}$ as the solution to the estimating equation 
\[
 \mathbb{P}_n\frac{2Z-1}{f(Z\mid X;\hat{\gamma})}\left\{Y-D\theta(X;\alpha)-\hat{E}(Y\mid Z=0,X)+\hat{E}(D\mid Z=0,X)\theta(X;\alpha)\right\}=0.
\]

\item[{\tt dru.simple}] If $\theta(X)=\LATE(X)$, then we assume
     \begin{flalign*}
P(Z=1\mid X) &= {\rm expit}(\gamma^\top X);\\
E(Y\mid X)&={\rm expit}(\varsigma^\top X);\\
E(D\mid X)&={\rm expit}(\vartheta^\top X).
\end{flalign*}
If $\theta(X)=\MLATE(X)$, then we assume 
 \begin{flalign*}
 P(Z=1\mid X) &= {\rm expit}(\gamma^\top X);\\
 E(Y \mid D,X)&={\rm expit}(\varpi_1 D+\varpi_2^\top X);\\
E(D\mid X)&={\rm expit}(\vartheta^\top X).
\end{flalign*}
In {\tt dru.simple}, first, we fit the models  $E(Y\mid X; \varsigma)$, $E(D\mid X; \vartheta)$, $E(Y \mid D,X; \varpi)$ and $P(Z=1\mid X;\gamma)$ using  {\tt R} package {\tt glm}.
Then, we obtain $\hat{\alpha}_{\rm dru.simple}$ as the solution to the estimating equation 
\[
 \mathbb{P}_n\frac{2Z-1}{f(Z\mid X;\hat{\gamma})}\left\{H(\alpha)-\hat{E}(H\mid X;\alpha)\right\}=0,
\]
where 
\[
\hat{E}\{H\mid X;\alpha\}=
\begin{cases}
E(Y\mid X;\hat{\varsigma})-E(D|X;\hat{\vartheta})\theta(X;\alpha) & \theta(X)=\LATE(X);\\
E(D\mid X;\hat{\vartheta})E(Y \mid D=1, X;\hat{\varpi})\theta(X;\alpha)^{-1}\\
\quad +\{1-E(D\mid X;\hat{\vartheta})\}E(Y\mid D=0,X;\hat{\varpi}) & \theta(X)=\MLATE(X).
\end{cases}
\]

\item[\tt ls.abadie] If $\theta(X)=\LATE(X)$, then we assume 
 \begin{flalign*}
 P(Z=1\mid X) &= {\rm expit}(\gamma^\top X);\\
E\{Y\mid X,D,D(1)>D(0)\}&=D \tanh(\alpha^\top X)+{\rm expit}(\varphi_1^\top X).
\end{flalign*}
If $\theta(X)=\MLATE(X)$, then we assume 
 \begin{flalign*}
 P(Z=1\mid X) &= {\rm expit}(\gamma^\top X);\\
E\{Y\mid X,D,D(1)>D(0)\}&=\{\exp(\alpha^\top X)\}^D {\rm expit}(\varphi_2^\top X).
\end{flalign*}
In {\tt ls.abadie}, first, we fit the model $P(Z=1\mid X;\gamma)$  using {\tt R} package {\tt glm}. Then, we obtain the weighted least squares estimator $(\hat{\alpha}_{\rm ls.abadie},\hat{\varphi}_{\rm ls.abadie})$ by minimizing the following objective function
\[
 \mathbb{P}_n w(X;\hat{\gamma})\left[Y-E\{Y\mid X,D,D(1)>D(0);\alpha,\varphi\}\right]^2,
\]
where 
\[
w(X;\hat{\gamma})=1-D(1-Z)/\{1-P(Z=1\mid X;\hat{\gamma})\}-(1-D)Z/P(Z=1\mid X;\hat{\gamma}).
\]

 \item[{\tt mle.crude}] If $\theta(X)=\LATE(X)$, then we assume
 \begin{flalign*}
E(Y\mid D=1,X)-E(Y\mid D=0,X)&=\tanh(\alpha^\top X).
 \end{flalign*}
 If $\theta(X)=\MLATE(X)$, then we assume
 \begin{flalign*}
\frac{E(Y\mid D=1,X)}{E(Y\mid D=0,X)}&=\exp(\alpha^\top X).
\end{flalign*}
In addition, for both cases, we assume 
 \begin{flalign*}
 {OP}^{YD}(X)&\equiv \frac{P(Y=1\mid D=1,X)P(Y=1\mid D=0,X)}{P(Y=0\mid D=1,X)P(Y=0\mid D=0,X)}=\exp(\rho^\top X);\\
 E(D \mid X)&={\rm expit}(\upsilon^\top X).
\end{flalign*}
 The above models are fitted using the maximum likelihood estimation method based on the likelihood function of $Y$ conditional on $D$ and $X$, and obtained using {\tt R} package {\tt brm}.  
\end{itemize}

\subsection{Detailed simulation results}

Table \ref{tab:biassd2} presents a detailed version of Table 2 in the main paper. Table \ref{tab:bias/sd} presents the ratio between Bias and SD for different estimators and scenarios considered in Table \ref{tab:biassd2}. Table \ref{tab:ci2} presents a detailed version of Table 3 in the main paper.

\subsection{The causal model assumed in the application to 401(k) data}

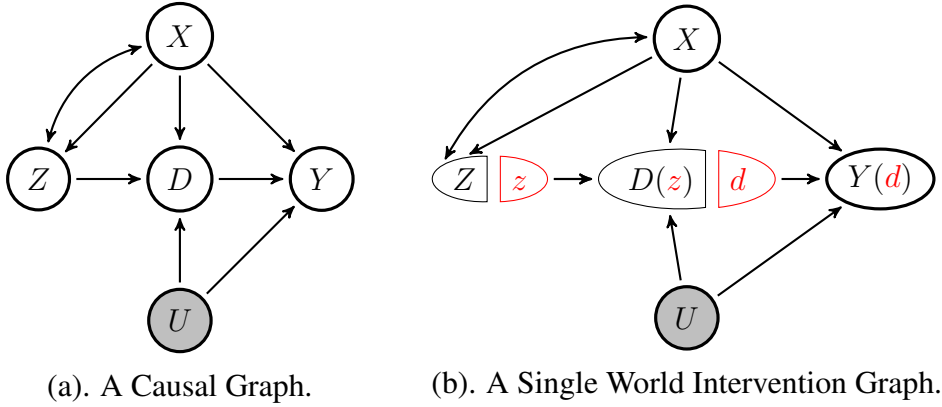
\begin{figure}[!htbp]
		\begin{tikzpicture}[->,>=stealth',node distance=1cm,pre/.style={->,>=stealth,ultra thick,black,line width = 1.5pt}, 
		swig vsplit={gap=5pt, line color right=red}]] %% Set style for split nodes
		%nodes
		\begin{scope}
		\node[est] (Z) {$Z$};
		\node[est, right = of Z] (D) {$D$};
		\node[est, right = of D] (Y) {$Y$};
		\node[shade, below = of D] (U) {$U$};
		\node[est, above = of D] (X) {$X$};
		\path[pil] (Z) edgenode {} (D);
		\path[pil] (D) edgenode {} (Y);
		\path[pil] (U) edgenode {} (D);
		\path[pil] (U) edgenode {} (Y);
		\path[pil] (X) edgenode {} (Z);
		\path[pil] (X) edgenode {} (D);
		\path[pil] (X) edgenode {} (Y);
		\path[pil, <->] (X) edge [bend right] node {} (Z);
		\node[name=la, below=2cm of D]{(a). A Causal Graph.};
		\end{scope}
%%%%%%%%%%%%%%%%%%%%%%%%%%%%%
		\begin{scope}[xshift=6cm]
		\node[name=Z, shape=swig vsplit]{ %Construct a vertically split "node"
			\nodepart[est]{left}{$Z$} % left half
			\nodepart[red]{right}{$z$}
			};
		\node[name=D, shape=swig vsplit, right=0.65cm of Z]{  %Construct a vertically split "node"
			\nodepart[est]{left}{$D({\color{red}{z}})$}  % left half
			\nodepart[red]{right}{$d$} % right half
			}; 
		\node[est, name=Y,shape=ellipse,style={draw}, right =0.65cm of D,inner sep=2pt]{$Y({\color{red}{d}})$};
		\draw[pil,->] (Z) to (D);
		\node[est, above =1cm of D] (X) {$X$};
		\path[pil] (Z) edgenode {} (D);
		\path[pil] (D) edgenode {} (Y);
		\draw[pil, ->] (X)  to  (Z.130);
		\path[pil, <->] (X) edge [bend right] node {} (Z.160);
		\draw[pil, ->] (X)  to  (D.110); %D.110 is angle 110
		\path[pil] (X) edgenode {} (Y);
		\node[shade, below = of D] (U) {$U$};
		\path[pil] (U) edgenode {} (D.250);% D.250 is angle 250
		\path[pil] (U) edgenode {} (Y);
		\node[name=la, below=2cm of D]{(b). A Single World Intervention Graph.};
		\end{scope}
		\end{tikzpicture}	
	\caption{The causal model assumed in the  application to 401(k) data. Variables $X,Z,D,Y$ are observed; $U$ is unobserved. The bi-directed edge between $X$ and $Z$  denotes unmeasured common cause.}
	\label{DAG:iv_model2}
\end{figure}

\subsection{Implementation details in the application to 401(k) data}
In the 401(k) data, since only eligible individuals may choose to participate in 401(k) plans, some model assumptions are different from those in the simulation studies. The model fitting methods are the same as that in  Section \ref{ss:imsimu}. In the application, we focus on the case where $\theta(X)=\MLATE(X)$. Models are fitted in a similar fashion as in Section \ref{ss:imsimu}.

\begin{itemize}
\item[\tt mle]
\item[\tt drw]
Assume
\begin{flalign*}
\phi_1(X)&={\rm expit}(\beta_1^\top X);\\ \phi_3(X)&={\rm expit}(\beta_2^\top X);\\
\quad OP^{CO}(X)&=\exp (\eta^\top X); \\
P(Z=1\mid {X}) &= {\rm expit}(\gamma^\top {X}).
\end{flalign*}
\item[\tt dru.ogburn]
\item[\tt drw.ogburn]
Assume 
\begin{flalign*}
P(Z=1\mid {X}) &= {\rm expit}(\gamma^\top {X});\\
E(H\mid X)&=E(Y\mid Z=0,X)={\rm expit}(\xi^\top X);\\
E(DY\mid Z=1,X)&={\rm expit}(\psi^\top X);\\
E \left[\left. \dfrac{\left\{H - E(H\mid X)\right\}^2}{f^2(Z\mid X)} \right| X \right]  &= \exp(\zeta^\top X).
\end{flalign*}

\item[{\tt dru.simple}] 
Assume
 \begin{flalign*}
 P(Z=1\mid {X}) &= {\rm expit}(\gamma^\top {X});\\
 E(Y \mid D,X)&={\rm expit}(\varpi_1 D+\varpi_2^\top X);\\
E(D\mid X)&={\rm expit}(\vartheta^\top X).
\end{flalign*}

\item[\tt ls.abadie]
Assume 
 \begin{flalign*}
 P(Z=1\mid {X}) &= {\rm expit}(\gamma^\top {X});\\
E\{Y\mid X,D,D(1)>D(0)\}&=\{\exp(\alpha^\top X)\}^D {\rm expit}(\varphi^\top X).
\end{flalign*}

 \item[{\tt mle.crude}] 
 Assume
 \begin{flalign*}
 \frac{E(Y\mid D=1,X)}{E(Y\mid D=0,X)}&=\exp(\alpha^\top X);\\
 {OP}^{YD}(X)&\equiv \frac{P(Y=1\mid D=1,X)P(Y=1\mid D=0,X)}{P(Y=0\mid D=1,X)P(Y=0\mid D=0,X)}=\exp(\rho^\top X);\\
 E(D \mid X)&={\rm expit}(\upsilon^\top X).
\end{flalign*}

\end{itemize}

	\newpage

\begin{table}[H]
	\centering
	\caption{The biases and standard errors of the estimated biases in the Monte-Carlo study of various estimators. The true value for $\alpha_0$ and $\alpha_1$ is 0 and -1, respectively. The sample size is  1000 }
	\bigskip
	\small
    \label{tab:biassd2}
	\begin{tabular}{rccccccc}
	\toprule
&       \multicolumn{2}{c}	{$\theta(X) = \LATE(X)$} &  \multicolumn{2}{c}	{ $\theta(X) = \MLATE(X)$} \\
\cmidrule(r){2-3} \cmidrule(l){4-5}
& \multicolumn{1}{c}{$\alpha_0$} & \multicolumn{1}{c}{$\alpha_1$} &   \multicolumn{1}{c}{$\alpha_0 $} & \multicolumn{1}{c}{$\alpha_1 $} \\
\midrule
\multicolumn{2}{l}{Bias $\times 100$ (SE $\times 100$) \quad \quad}	  & & \\ [5pt]
	{\quad \quad}	mle.bth & 0.28(0.35) & -3.5(0.78) & -0.092(0.71) & -3.0(1.2) \\ 
mle.bad & -20(0.42) & -15(0.80) & -48(1.2) & -18(2.1) \\ [3pt]
drw.bth & 0.55(0.36) & -4.1(0.82) & 0.54(0.77) & -5.6(1.5) \\ 
drw.psc & 0.060(0.38) & -5.9(1.0) & -0.38(1.2) & -12(2.7) \\ 
drw.opc & 0.55(0.36) & -3.9(0.79) & 0.49(0.75) & -5.3(1.4) \\ 
drw.bad & -10(0.40) & -9.6(1.1) & -28(1.4) & 25(3.3)  \\[3pt] 
dru.bth & 1.3(0.44) & -5.8(1.0) & 1.8(0.84) & -8.1(1.7) \\ 
dru.psc & 1.2(0.44) & -6.1(1.0) & 1.9(0.84) & -9.0(1.7) \\ 
dru.opc & 0.94(0.39) & -4.5(0.86) & 0.99(0.78) & -6.5(1.5) \\ 
dru.bad & -14(0.48) & -27(1.4) & -28(0.98) & -12(2.4) \\ [5pt]
reg.ogburn.bth & -5.7(1.6) & -2.9(3.1) & 7.8(2.0) & -1.1(2.2) \\ 
reg.ogburn.bad & -9.0(0.25) & 100(0.23) & 140(5.6) & 93(3.6) \\ [3pt] 
drw.ogburn.bth & 0.10(0.46) & -4.2(0.99) & 3.2(1.4) & -13(2.5) \\ 
drw.ogburn.psc & 1.3(0.46) & -8.2(1.3) & 7.7(1.9) & -18(3.5) \\ 
drw.ogburn.opc & -5.5(1.1) & -8.0(1.5) & 12(2.3) & 21(4.7) \\ 
drw.ogburn.bad & -120(3.1) & -170(6.0) & -40(2.6) & -9.4(5.8) \\  [3pt]
dru.ogburn.bth & 1.3(0.45) & -5.8(1.1) & 1.9(0.85) & -8.2(1.7) \\ 
dru.ogburn.psc & 1.5(0.49) & -9.1(1.3) & 3.1(0.90) & -11(2.0) \\ 
dru.ogburn.opc & -2.9(0.58) & -2.5(1.1) & 4.6(1.5) & 11(3.0) \\ 
dru.ogburn.bad & -130(3.2) & -190(6.4) & -47(1.0) & -41(3.7) \\   [5pt]
mle.wang.bth & 0.28(0.35) & -3.8(0.79) & $-$ & $-$ \\ 
mle.wang.bad & -27(0.40) & -1.5(1.1) & $-$ & $-$ \\ [3pt]
dru.wang.bth & 1.3(0.45) & -5.8(1.0) & $-$ & $-$ \\  
dru.wang.psc & 1.2(0.45) & -6.3(1.0) & $-$ & $-$ \\ 
dru.wang.opc & 0.29(0.41) & -7.8(1.0) &  $-$ & $-$ \\  
dru.wang.bad & -20(0.52) & -24(1.5) & $-$ & $-$ \\  [5pt]
dru.simple.bth & 1.3(0.45) & -5.8(1.0) & 1.8(0.84) & -8.0(1.7) \\ 
dru.simple.psc & 1.2(0.44) & -6.2(1.0) &1.9(0.84) & -8.8(1.7) \\ 
dru.simple.opc & 4.5(0.49) & -17(1.2) & -0.15(0.68) & 11(1.2) \\ 
dru.simple.bad & -16(0.48) & -17(1.3) & -34(0.70) & 18(1.5) \\ [5pt] 
ls.abadie.bth & -0.19(0.37) & -4.1(0.93) & 0.42(0.79) & -11(1.6) \\ 
ls.abadie.bad & -23(0.88) & 22(1.2) & -32(1.9) & 7.7(3.6) \\ [5pt] 
mle.crude & -2.8(0.10) & 60(0.19) & 0.36(0.25) & 51(0.42) \\ 
			\bottomrule 
	\end{tabular}
\end{table}

\begin{table}[H]
	\centering
	\caption{The Bias/SD of all the estimators. The true values for $\alpha_0$ and $\alpha_1$ are 0 and -1, respectively. The sample size is  1000 }
	\bigskip
	\small
    \label{tab:bias/sd}
	\begin{tabular}{rrrrrc}
	\toprule
&       \multicolumn{2}{c}	{$\theta(X) = \LATE(X)$} &  \multicolumn{2}{c}	{ $\theta(X) = \MLATE(X)$} \\
\cmidrule(r){2-3} \cmidrule(l){4-5}
& \multicolumn{1}{c}{$\alpha_0$} & \multicolumn{1}{c}{$\alpha_1$} &   \multicolumn{1}{c}{$\alpha_0 $} & \multicolumn{1}{c}{$\alpha_1 $} \\
\midrule
	\multicolumn{2}{l}{Bias/SD \quad \quad}	 &  & \\ [5pt]
	{\quad \quad}	mle.bth & 0.02 & -0.14 & -0.00 & -0.08 \\ 
mle.bad & -1.53 & -0.61 & -1.25 & -0.28 \\ [3pt]
drw.bth & 0.05 & -0.16 & 0.02 & -0.12 \\ 
drw.psc & 0.01 & -0.19 & -0.01 & -0.14 \\ 
drw.opc & 0.05 & -0.15 & 0.02 & -0.12 \\ 
drw.bad & -0.79 & -0.28 & -0.66 & 0.24 \\ [3pt]
dru.bth & 0.09 & -0.18 & 0.07 & -0.15 \\ 
dru.psc & 0.08 & -0.19 & 0.07 & -0.16 \\ 
dru.opc & 0.08 & -0.17 & 0.04 & -0.13 \\ 
dru.bad & -0.94 & -0.61 & -0.89 & -0.15 \\ [5pt]
reg.ogburn.bth & -0.11 & -0.03 & 0.13 & -0.02 \\ 
reg.ogburn.bad & -1.16 & 14.18 & 0.81 & 0.82 \\ [3pt]
drw.ogburn.bth & 0.01 & -0.13 & 0.07 & -0.16 \\ 
drw.ogburn.psc & 0.09 & -0.20 & 0.13 & -0.16 \\ 
drw.ogburn.opc & -0.15 & -0.17 & 0.17 & 0.14 \\ 
drw.ogburn.bad & -1.48 & -1.07 & -0.49 & -0.05 \\ [3pt]
dru.ogburn.bth & 0.09 & -0.17 & 0.07 & -0.15 \\ 
dru.ogburn.psc & 0.10 & -0.22 & 0.11 & -0.17 \\ 
dru.ogburn.opc & -0.16 & -0.07 & 0.10 & 0.12 \\ 
dru.ogburn.bad & -1.53 & -1.10 & -1.47 & -0.35 \\[5pt]
mle.wang.bth & 0.03 & -0.15 & $-$ & $-$ \\
mle.wang.bad & -2.16 & -0.04 & $-$ & $-$ \\ [3pt]
dru.wang.bth & 0.09 & -0.18 & $-$ & $-$ \\
dru.wang.psc & 0.09 & -0.19 & $-$ & $-$ \\
dru.wang.opc & 0.02 & -0.25 & $-$ & $-$ \\
dru.wang.bad & -1.24 & -0.52 & $-$ & $-$ \\ [5pt]
dru.simple.bth & 0.09 & -0.18 & 0.07 & -0.15 \\ 
dru.simple.psc & 0.09 & -0.19 &0.07 & -0.16 \\ 
dru.simple.opc & 0.29 & -0.44 & -0.01 & 0.28 \\ 
dru.simple.bad & -1.07 & -0.41 &  -1.54 & 0.36 \\[5pt]
ls.abadie.bth & -0.02 & -0.14 & 0.02 & -0.22 \\ 
ls.abadie.bad & -0.84 & 0.57 & -0.54 & 0.07 \\ [5pt]
mle.crude & -0.83 & 10.02 & 0.04 & 3.76 \\ 
			\bottomrule 
	\end{tabular}
\end{table}

\begin{table}[H]
	\centering
	\caption{The coverage probabilities and average widths of confidence intervals obtained from $500$ bootstrap samples. The true values for $\alpha_0$ and $\alpha_1$ are 0 and -1, respectively. The sample size is  1000}
	\bigskip
	\small
    \label{tab:ci2}
	\begin{tabular}{rcccccc}
	\toprule
&       \multicolumn{2}{c}	{$\theta(X) = \LATE(X)$} &  \multicolumn{2}{c}	{ $\theta(X) = \MLATE(X)$} \\
\cmidrule(r){2-3} \cmidrule(l){4-5}
& \multicolumn{1}{c}{$\alpha_0$} & \multicolumn{1}{c}{$\alpha_1$} &   \multicolumn{1}{c}{$\alpha_0 $} & \multicolumn{1}{c}{$\alpha_1 $} \\
\midrule
\multicolumn{2}{l}{Coverage probability $\times 100$ \quad \quad}	  & & \\ [5pt]
	{\quad \quad}	mle.bth & 95.6(0.502) & 95.8(1.25) & 95.4(1.12) & 96.4(2.02) \\ 
mle.bad & 65.4(0.662) & 91.0(1.38) & 46.3(2.00) & 94.6(3.38) \\ [3pt]
drw.bth & 94.7(0.513) & 95.2(1.31) & 96.9(1.21) & 95.9(2.34) \\ 
drw.psc & 95.4(0.561) & 95.5(1.65) & 97.6(2.43) & 97.4(4.20) \\ 
drw.opc & 95.0(0.505) & 95.6(1.25) & 96.1(1.19) & 96.1(2.30) \\ 
drw.bad & 87.0(0.614) & 95.3(1.82) & 91.8(2.94) & 96.8(5.54) \\[3pt] 
dru.bth & 94.5(0.636) & 94.6(1.59) & 96.3(1.46) & 96.9(2.61) \\ 
dru.psc & 94.8(0.639) & 95.0(1.62) & 96.3(1.46) & 97.6(2.65) \\ 
dru.opc & 93.8(0.564) & 94.8(1.34) & 96.1(1.31) & 96.9(2.33) \\ 
dru.bad & 79.6(0.732) & 91.3(2.35) & 87.4(1.61) & 98.2(3.28) \\ [5pt]
reg.ogburn.bth & 98.0(1.37) & 99.9(3.40) & 99.6(2.38) & 100.0(3.09) \\ 
reg.ogburn.bad & 75.6(0.318) & 0.1(0.284) & 99.9(4.81) & 86.1(3.78) \\ [3pt] 
drw.ogburn.bth & 97.0(0.564) & 98.1(1.42) & 98.5(1.85) & 98.4(3.29) \\ 
drw.ogburn.psc & 95.4(0.689) & 95.7(2.01) & 99.7(2.95) & 99.9(5.18) \\ 
drw.ogburn.opc & 98.8(0.830) & 99.0(1.98) & 99.8(2.39) & 99.9(5.14) \\ 
drw.ogburn.bad & 2.2(2.84) & 72.5(5.62) & 99.7(3.06) & 99.9(6.55) \\ [3pt] 
dru.ogburn.bth & 94.5(0.644) & 95.0(1.63) & 96.2(1.37) & 97.2(2.59) \\ 
dru.ogburn.psc & 93.4(0.692) & 94.3(2.03) & 96.9(1.51) & 97.3(2.85) \\ 
dru.ogburn.opc & 99.3(0.836) & 97.9(1.75) & 99.6(1.76) & 99.6(3.79) \\ 
dru.ogburn.bad & 1.6(2.95) & 64.0(5.81) & 91.6(1.39) & 99.6(4.27) \\ [5pt] 
mle.wang.bth & 94.3(0.466) & 95.1(1.13) & $-$ &  $-$ \\ 
mle.wang.bad & 41.1(0.564) & 94.3(1.41) & $-$ &  $-$ \\  [3pt]
dru.wang.bth & 94.4(0.637) & 94.7(1.60) &  $-$ &  $-$  \\ 
dru.wang.psc & 94.8(0.643) & 94.7(1.65) &  $-$ &  $-$  \\ 
dru.wang.opc & 94.1(0.581) & 94.5(1.60) &  $-$ &  $-$  \\ 
dru.wang.bad & 68.0(0.786) & 92.1(2.47) &  $-$ &  $-$  \\ [5pt]
dru.simple.bth & 94.3(0.636) & 94.8(1.59) & 96.1(1.36) & 96.5(2.51) \\ 
dru.simple.psc & 94.6(0.641) & 94.5(1.63) & 96.1(1.38) & 96.8(2.57) \\ 
dru.simple.opc & 93.6(0.692) & 92.5(1.84) & 96.2(1.05) & 94.9(1.79) \\ 
dru.simple.bad & 76.4(0.706) & 94.2(2.16) & 69.3(1.05) & 93.8(2.27) \\ [5pt]
ls.abadie.bth & 94.8(0.593) & 95.5(1.57) & 96.4(1.24) & 95.6(2.56) \\ 
ls.abadie.bad & 87.3(0.898) & 94.5(1.50) & 93.1(2.03) & 95.7(3.86) \\ [5pt]
mle.crude & 84.3(0.127) & 0.0(0.228) & 94.0(0.303) & 6.3(0.524) \\ 
			\bottomrule 
	\end{tabular}
\end{table}

\end{document}